\documentclass[prd,preprint,superscriptaddress,nofootinbib,longbibliography]{revtex4}

\newcommand{\be}{\begin{equation}}
	\newcommand{\ee}{\end{equation}}
\def\lta{\,\raise 0.3 ex\hbox{$ < $}\kern -0.75 em
	\lower 0.7 ex\hbox{$\sim$}\,}
\def\gta{\,\raise 0.3 ex\hbox{$ > $}\kern -0.75 em
	\lower 0.7 ex\hbox{$\sim$}\,} 
\newcommand{\amp}{{\bar\Phi}}

\newcommand{\mpl}{M_{\rm pl}}
\newcommand{\phire}{\phi_{\scriptstyle R}}
\newcommand{\tred}{t_{\scriptstyle R}}

\usepackage{xcolor}

\usepackage{graphicx}
\usepackage{dcolumn}
\usepackage{bm,amsfonts,amsthm,amsmath,amssymb,array}
\usepackage{bbm}
\usepackage[]{epsfig,graphics}
\usepackage{comment}
\usepackage[T1]{fontenc}
\usepackage{lipsum}
\usepackage[utf8]{inputenc}
\usepackage{ulem}
\usepackage{multirow}
\usepackage{color}
\usepackage{lastpage}
\usepackage[sort&compress]{natbib}
\usepackage{enumerate}
\usepackage{wasysym}
\usepackage{relsize}
\usepackage{physics}
\usepackage{mathrsfs}
\usepackage{tensor}
\usepackage{xfrac}
\usepackage{siunitx}
\usepackage{setspace}  
\usepackage[
colorlinks=true,     
urlcolor=blue,       
]{hyperref}
\usepackage{caption}
\usepackage{subcaption}
\def\ben{\begin{enumerate*}}
	\def\een{\end{enumerate*}}
\def\bi{\begin{itemize*}}
	\def\ei{\end{itemize*}}
\def\bd{\begin{description*}}
	\def\ed{\end{description*}}
\def\be{\begin{equation}}
	\def\ee{\end{equation}}
\def\bea{\begin{eqnarray}}
	\def\eea{\end{eqnarray}}
\def\bfl{\begin{flushleft}}
	\def\efl{\end{flushleft}}

\newcommand{\frw}{{\mbox{\tiny FRW}}}
\newcommand{\osc}{{\mbox{\tiny osc}}}

\newcommand{\gsim}{\lower.7ex\hbox{$\;\stackrel{\textstyle>}{\sim}\;$}}
\newcommand{\lsim}{\lower.7ex\hbox{$\;\stackrel{\textstyle<}{\sim}\;$}}

\newcommand{\beq}{\begin{equation}}
	\newcommand{\eeq}{\end{equation}}

\begin{document} 
	
	\title{Generalized Models for Inflationary Preheating:\\ Oscillations and Symmetries}

	
	\author{Leia Barrowes}
	\email{barrowes@umich.edu}
	\affiliation{Physics Department, University of Michigan, Ann Arbor, MI 48109, USA} 
	
	\author{Fred C. Adams}
	\email{fca@umich.edu}
	\affiliation{Physics Department, University of Michigan, Ann Arbor, MI 48109, USA} 
	\affiliation{Astronomy Department, University of Michigan, Ann Arbor, MI 48109, USA}
	\author{Anthony M. Bloch}
	\email{abloch@umich.edu}
	\affiliation{Mathematics Department, University of Michigan, Ann Arbor, MI 48109, USA}
	
	\author{John T. Giblin, Jr.}
	\email{giblinj@kenyon.edu}
	\affiliation{Department of Physics, Kenyon College, Gambier, Ohio 43022, U.S.A.}
	\affiliation{Department of Physics, Case Western Reserve University, Cleveland, Ohio 44106, U.S.A.}
	\affiliation{Center for Cosmology and AstroParticle Physics (CCAPP) and Department of Physics, Ohio State University, Columbus, OH 43210, USA}
	
	\author{Scott Watson}
	\email{gswatson@syr.edu}
	\affiliation{Department of Physics, Syracuse
		University, Syracuse, NY 13244, USA} 
	\affiliation{Department of Physics and Astronomy, University of South Carolina, Columbia, SC 29208, USA}
	
	\begin{abstract}
		The paradigm of the inflationary universe provides a possible explanation for several observed cosmological properties. In order for such solutions to be successful, the universe must convert the energy stored in the inflaton potential into standard model particles through a process known as reheating.  In this paper, we reconsider the reheating process for the case where the inflaton potential respects an approximate (but spontaneously broken) conformal symmetry during the reheating epoch. After reviewing the Effective Field Theory of Reheating, we present solutions for the nonlinear oscillations of the inflaton field, derive the corresponding Hill's equation for the coupled reheating field, and determine the stability diagram for parametric resonance. For this class of models ---the simplest realization being a scalar field with a quartic term---the expansion of the universe drives the coupled field toward a more unstable part of parameter space, in contrast to the standard case.
		We also generalize this class of models to include quadratic breaking terms in the potential during the reheating epoch and address the process of stability in that universality class of models. 
	\end{abstract}
	
	\maketitle
	
	\section{Introduction}
	\label{sec:intro}
	
	Many observed properties of our universe can be understood if the universe went through an epoch of exponential expansion early in cosmic history \cite{Guth1981}. This inflationary phase helps explain the homogeneity and flatness of the universe, along with its apparent lack of magnetic monopoles. In addition, fluctuations in the energy density of the inflationary field can be identified with the perturbations observed in the cosmic microwave background. At the end of the inflationary epoch, the universe finds itself in a state where energy is primarily stored in the potential of the inflaton field, the scalar field that was driving the exponential expansion. In order to transition from this inflationary phase to a radiation dominated era, and thereby recover the standard description of Big Bang cosmology, a crucial process known as reheating must occur. Reheating is responsible for converting the potential energy of the inflaton field into the standard model particles that make up our observable universe. The goal of this paper is to explore a particular scenario for the reheating phase, where the oscillations of the inflaton field are driven by a quartic potential. As outlined below, although the inflaton potential in this scenario has a quartic form during reheating,  it can have a different form earlier during the slow-roll phase. Specifically, we examine the prospects for parametric resonance within this quartic model and show that it helps facilitate successful reheating. 
	
	Since a large amount of previous work on reheating has been carried out (see the reviews of \cite{Amin:2014eta,Lozanov2019}), we start with a brief overview. First, it is important to differentiate between the processes of reheating and preheating. As considered here, reheating is the general process of converting the potential energy of the inflaton field into radiation and matter, thereby establishing the hot and dense conditions required for the early universe. In contrast, preheating refers to the specific mechanism of energy conversion driven by parametric resonance that arises in particular inflationary models. During the oscillatory phase of the inflaton field around the minimum of its potential, the inflaton couples to additional fields that in turn couple to standard model particles. If these additional fields experience the phenomenon of parametric resonance, which results in rapid and explosive particle production, then efficient energy conversion can occur.   
	
	This paper considers a particular preheating scenario, where the inflaton field features a quartic potential during its oscillatory phase of preheating, building on the seminal work of \cite{Greene_1997}. The quartic potential results in nonlinear oscillations of the inflaton field. These nonlinearities, in turn, result in different forms for the variables that drive parametric resonance. More specifically, while parametric resonance is often described by the Mathieu equation \cite{Mathieu1868}, this scenario leads to a more general Hill's equation \cite{Hill1886,MagWink1966}, which has different bands of instability (and stability) for resonance. More significantly, the effective forcing parameter in Hill's equation grows with continued expansion of the universe, thus leading to more instability. This behavior stands in contrast to most previously considered cases where the evolution of the parameters in the Mathieu equation leads to greater stability.  
	We build upon the methods of \cite{Greene_1997} and show the Floquet maps for a wide range of scenarios with a primary focus on the efficient trilinear interaction, and determine the trajectories of the system through parameter space.
	In addition, we include numerical results to support our semianalytic work.
	
	While the scenario of this paper considers the inflaton field to have a quartic form during the reheating epoch, the potential can have a more general form at earlier times \citep{Ozsoy:2017mqc}.  During the part of inflation when the universe expands exponentially, usually operating under slow-roll conditions, a quartic form for the potential is currently disfavored by observations of the cosmic microwave background. These observations constrain the form of the inflaton potential for times corresponding to 70 to 60 e-foldings before the end of the slow-roll era, whereas the reheating epoch takes place after its conclusion.
	
	This paper is organized as follows. We first discuss (in Section \ref{sec:fieldtheory}) how Effective Field Theory (EFT) can lead to different forms for the inflaton potential during the slow-roll epoch (when observable density fluctuations are produced) and the subsequent reheating epoch. Next we solve for the nonlinear oscillations of the inflaton field for the case of a quartic potential (in Section \ref{sec:model}), and derive the corresponding Hill's equation for reheating. We then carry out a Floquet analysis of the resulting model equation and find its stability diagram (in Section \ref{sec:parametric}).  As outlined above, the expansion of the universe drives the forcing parameter in Hill's equation to larger values, leading to greater instability. If the inflaton potential has a mass term as well as a quartic term during reheating, then the behavior of the reheating field is more complicated, and we address this issue in Section \ref{sec:massterm}. The paper concludes in Section \ref{sec:conclude} with a summary of results and a discussion of their implications.  
	
	\section{Challenges for Reheating and the EFT Approach}
	\label{sec:fieldtheory}
	
	Before considering parametric resonance for specific preheating scenarios (see the following section), it is useful to place this work in context. Methods to describe symmetries and symmetry breaking are a cornerstone of physics, and the transition from an inflationary epoch to the reheating of the universe can be described using this methodology. The cosmic expansion history is generally characterized by the scale factor and Hubble parameter, but these quantities are only well-defined for a perfectly homogeneous and isotropic universe.
	As inflation comes to an end, the cosmic expansion changes, and the cosmology can become difficult to analyze. Many expansion histories are possible. Particles are created in inhomogeneous regions. The physical mechanisms that drive  reheating and recover the standard `Hot Big Bang' model involve
	parametric resonance, turbulence, and chaotic behavior, all of which occur prior to thermalization. Moreover, most previous efforts to understand the reheating process have been highly model dependent. 
	
	An alternative approach to understanding the dynamics of reheating is to use methods of Effective Field Theory (EFT) to describe the process \cite{Ozsoy:2017mqc,Ozsoy:2015rna}. Such methods have been useful in particle physics \cite{Petrov:2016azi,Penco:2020kvy}, dark energy \cite{Park:2010cw,Bloomfield:2012ff,Gubitosi:2012hu,Wen:2021bsc} and condensed matter physics \cite{Brauner:2022rvf}. In the present context, the crucial feature of the EFT is that it allows us to establish universality classes  \cite{Ozsoy:2017mqc,Ozsoy:2015rna}. The cosmic expansion for a given equation of state is determined, but the dynamics and possible departures can be classified in symmetry groups, which define the universality classes. In the case of dark energy, for example, it has been shown that different ways of triggering cosmic acceleration come in such classes respecting their common symmetries
	\cite{Bloomfield:2012ff,Gubitosi:2012hu,Wen:2021bsc}.
	What we would like to emphasize in this paper is that symmetries could be a broader mechanism for discovery the dynamics of inflation reheating. {\bf The inflaton need not be a scalar field.} In addition, it is important to emphasize the work of \cite{Greene:2000ew} which first demonstrated that fermion preheating is also possible and efficient\footnote{We thank our journal referee for reminding of this important point, and it indeed was part of our motivation for broadening our approach to the notion of (p)reheating to be more general than simple scalar fields and their interactions.}.  Instead, it can be a `clock' or spurion of broken time-translation invariance. As we learn more about the nature of inflation this creates a useful framework, or methodology for exploring (p)reheating. 
	
	This paper focuses on conformal symmetry and its subsequent breaking during the epochs following inflation\footnote{In the case of conformal symmetry it is important to note that although we are taking a different approach to the reasoning, there has been significant past related work, in particular \cite{Ozsoy:2015rna,Ozsoy:2017mqc}. The point here is that the EFT is a different way of thinking about the end of inflation, and we have also tried to present further results beyond the initial use of this method.}. Given that we do not know how inflation ends---and whether inflation is driven by a true scalar field---we approach this problem by considering the inflaton to be an order parameter, more specifically a clock that depends on time and but not on space. As a result, the inflaton breaks time-translation invariance (more precisely time diffeomorphism invariance), but it need not be a true scalar field. One can think of the field both as a clock and as a Goldstone boson of the broken time translation invariance. When considering  cosmological solutions, here an FLRW universe, the cosmic expansion breaks the symmetry in time. This theoretical notion has (previously) led to the EFT description of inflation and can be continued for the case of reheating. 
	
	During reheating, two time scales are crucially  important. The first is the expansion time, the inverse of the Hubble parameter. The second is the inflaton's oscillation time scale, where this period must be much shorter than the expansion time. Given this hierarchy, it is appropriate to use EFT methods for decoupling the time scales. We thus review the EFT approach and  consider the universality class of models with conformal symmetry, along with its subsequent breaking. 
	
	The EFT of reheating and structure formation is based on the idea that there is a physical clock corresponding to the Goldstone boson that non-linearly realizes the spontaneously broken time invariance of the background \cite{Ozsoy:2017mqc,Ozsoy:2015rna}.  In unitary gauge, where the clock is homogeneous, the matter perturbations are encoded within the metric, i.e., the would-be Goldstone bosons are `eaten' by the metric since  gravity is a gauge theory.   
	
	During reheating the inflaton undergoes oscillations, while quanta of the reheating fields are produced in a process that can be both complicated and highly model-dependent. The EFT approach utilizes the notion that as long as the inflaton dominates the energy density of the background universe, the expansion can be described through the ansatz  
	\be \label{H_osc}
	H(t) = H_{\frw}(t) + H_{\osc}(t) P(\omega t),
	\ee
	where $H_{\frw}(t)$ is the overall (averaged) Hubble expansion rate. The second term leads to an oscillatory correction that is sub-dominant, so that $H_\frw \gg H_{\osc} P(\omega t)$, where $P(\omega t)$ is a quasi-periodic function. This evolution of the background spontaneously breaks time translations $t \rightarrow t + \xi^0(t,\vec{x})$, first to a discrete symmetry and then completely. As a result, if we probe the background at energies $E \gg H(t)$, corresponding to frequencies $\omega \gg H(t)$, then time symmetry is restored until we consider energies comparable to the frequency $\omega$. At such energies $H_{\frw}$ and $H_{\osc}$ will remain nearly constant and the time symmetry is nearly restored, but the symmetry will be broken by  $P(\omega t)$ to a discrete symmetry $t \rightarrow t + 2\pi \omega^{-1}$. At lower energies, both $H_{\frw}$ and $H_{\osc}$ will also evolve, thereby breaking the discrete symmetry. This case of symmetry breaking is a natural consequence of the hierarchy of scales that appears in reheating, i.e., high energy (short wavelength) modes probe inflaton oscillations $E / \omega$, whereas low energy (long wavelength) modes capture the expansion of the background $E / H_\frw$. We thus require $H_\frw / \omega \ll 1$ during reheating. These results were used in previous work \cite{Ozsoy:2017mqc,Ozsoy:2015rna}
	to construct the EFT of reheating in terms of the Goldstone boson that non-linearly realizes the time symmetry
	
	
	Before proceeding, it is useful to more fully elucidate the meaning of equation \eqref{H_osc}. For example, consider a simple reheating model where the inflaton oscillates in a potential $V \simeq m^2 \phi_0^2$. In this case, one can solve for the background evolution \cite{Ozsoy:2017mqc,Ozsoy:2015rna} and find that 
	\be \label{s1}
	H(t)=H_m - \frac{3H_m^2}{4m} \sin (mt+\delta) + \ldots, 
	\ee
	where $H_m = 2/(3t)$ is the Hubble parameter for a matter dominated universe with scale factor $a(t) \sim t^{2/3}$, $\delta$ is a constant phase, and the dots represent terms suppressed by higher powers of $H_m/m$. We see that equation \eqref{s1} is of the form of equation \eqref{H_osc}, corresponding to a matter dominated universe corrected by oscillations suppressed by powers of $H_m/m$. At energies comparable to the mass of the inflaton, the inflaton oscillations break the time symmetry, whereas for energies $H \lesssim E \ll m$ the matter dominated expansion itself is primarily responsible for the breaking. This result is familiar: On scales comparable to the Hubble radius, reheating with a massive inflaton oscillating in a quadratic potential looks like a matter dominated universe, whereas on small scales one can treat the particle production as a local process and in many cases neglect the presence of gravity. In general, however, the potential can be more complicated (not purely quadratic, as we consider in this work) and the expansion of the universe of large scales can depart from the $a\sim t^{2/3}$ behavior of matter domination.  
	
	Working with the EFT approach, we focus on a conformal class of models, and emphasize the interpretation of the inflaton as an order parameter. We also emphasize (again) that the form of the inflationary potential during reheating is not necessarily the same as during inflation, so that there is no observational motivation to dismiss quartic terms.  Such a potential results in an approximate conformal symmetry and produces interesting dynamics. If the inflation were strictly a scalar field, then an immediate concern would be Coleman-Weinberg corrections to the potential. Motivated by our ignorance of UV complete reheating, however, we assume that such a model is possible. For completeness, we also subsequently consider the presence of an additional mass term.
	
	
	The EFT approach removes the model dependence by focusing on a few properties shared by all such models. First, the hierarchy $\omega \gg H$ is preserved by all reasonable choices of potential. Second, all we need to construct the EFT is spontaneously broken time symmetry of the background evolution.  As stressed in \cite{Cheung:2007st}, the background itself is not an observable, but instead the perturbations about the background are observable.  Before proceeding to our results for a particular class of conformal models, we review the EFT of inflationary reheating.

	\subsection{Review of the EFT of Reheating}
	
	During reheating the energy density will evolve from inflaton oscillations into a relativistic bath of particles. The fields responsible for spontaneously breaking the time translation invariance will thus change. In \cite{Creminelli:2006xe} it was argued that the Goldstone approach holds for any FLRW universe \eqref{H_osc} and any number of matter fields $\phi_m(t)$ contributing to the background energy density with perturbation
	\be
	\delta \phi_m(t,\vec{x})= \phi_m\left(t+\pi(t,\vec{x})\right) - \phi_m(t).
	\ee
	This shift in the time in the long wavelength limit corresponds to the {\it adiabatic mode}, which Weinberg demonstrated obeys a conservation theorem regardless of the matter content of the universe \cite{Weinberg:2003sw}.  The field $\pi(\vec{x})$ is the desired Goldstone mode used in constructing the EFT.  
	We use $\pi(t,\vec{x})$ to construct our theory of reheating, noting that as inflatons are converted into reheating fields we are simply making use of the adiabatic mode description.   
	
	The procedure for constructing the EFT follows analogously to that for inflation \cite{Cheung:2007st}.
	Working in unitary gauge, the EFT of fluctuations for reheating in the gravitational and inflationary sectors is given by the action\footnote{We work in reduced Planck units $m_p=1/\sqrt{8 \pi G}=2.44 \times 10^{18}$ GeV with $\hbar=c=1$ and with a mostly plus $(-,+,+,+)$ sign convention for the metric.}
	{\small
		\be
		S=\int d^4x \sqrt{-g}\, \left[ \frac{1}{2} m_p^2 R + m_p^2 \dot{H}g^{00} - m_p^2 \left(3H^2+\dot{H} \right) \right. 
		+\left. \frac{M_2^4(t)}{2!} \left( \delta g^{00} \right)^2 +
		\frac{M_3^4(t)}{3!} \left( \delta g^{00} \right)^3 + \ldots \right],
		\label{theaction}
		\ee
		where $g^{00}=-1+\delta g^{00}$ and the dots represent terms higher order in fluctuations and derivatives.  
		Just as in the inflationary case one introduces the Goldstone boson $\pi$, which nonlinearly realizes broken time symmetry.  This ansatz forces nontrivial relations between the operators in the action of equation \eqref{theaction}, e.g., the parameter $M_2$ simultaneously modifies the speed of sound, as well as additional interactions.  Note that because of the symmetry breaking, the speed of sound in the presence of inflaton oscillations is not protected, so that in general $c_\pi \neq 1$. 
		
		It is useful to consider the decoupling limit where $\dot{H} \rightarrow 0$ and $m^2_p \rightarrow \infty$, while their product remains fixed.
		This limit makes more precise the usual assumption in (p)reheating that particle creation is a local process and one typically ignores contributions coming from gravitational terms. One then focuses on operators fixed by tadpole cancellation and take $M_2=M_3=\ldots=0$.  In spatially flat gauge, the quadratic action in the decoupling limit is
		{\small
			\be \label{decoupled_action}
			S^{(2)}=\int d^4x \,a^3 m_p^2 \left[ -\dot{H}\left(\dot{\pi}^2-a^{-2}(\partial_i \pi )^2 \right) -3\dot{H}^2 \pi^2 \right],
			\ee
		}which by canceling the tadpoles has left us with coefficients fixed by the background evolution.
		Introducing the canonical field ${\pi}_c =m_p (-2 \dot{H})^{1/2} {\pi}$ one can show that rather than mixing with gravity, the oscillatory behavior of the time-dependent potential of the inflaton (corresponding to an operator $\hat{{\cal O}}_\pi \sim V^{\prime \prime} \pi_c^2$), breaks the shift symmetry.
		As in the EFT of Inflation the leading mixing with gravity scales as $E_{\mbox{\tiny mix}}=\epsilon^{1/2} H=\dot{H}^{1/2}$, although a difference for us is given $V\sim \phi_0^n$ then $\epsilon=3n/(n+2)$ is typically a dimensionless number of order unity. The decoupling limit will be useful for probing scales with $E \gg E_{\mbox{\tiny mix}}$, but at other times it is appropriate to include corrections coming from the mixing with gravity.
		One useful aspect of \eqref{decoupled_action} is to study the stability of sub-horizon perturbations against collapse. Just as in studies of ghost condensation \cite{ArkaniHamed:2003uy}, including higher corrections to the EFT (e.g., $M_2 \neq 0$) could lead to new and consistent models for (p)reheating.
		
		Whether or not to take the decoupling limit depends on the situation. For example, in considering the behavior of modes that re-enter the horizon during reheating, it is useful to calculate the leading corrections to  \eqref{decoupled_action} coming from the mixing with gravity. In that case, we consider modes between the two hierarchical scales $k/a \ll m$ while $k/a \gtrsim H$.
		The leading order mixing term is then given by 
		\be
		\Delta S^{(2)} = -\frac{1}{2} \int d^4x \, a^3 \left( \frac{2 \ddot{H}}{H}\right) \pi_c^2,
		\ee
		which is written in terms of the canonical field $\pi_c$ \cite{Ozsoy:2017mqc,Ozsoy:2015rna}. One can show that this leading correction results in a growing, oscillatory contribution to the power spectrum. For a potential dominated by a mass term $V\sim m^2 \phi_0^2$, this correction leads to the main result of \cite{Easther:2010mr}, where those authors performed a full analysis, including all gravitational perturbations.
		
		Part of the utility of the EFT approach is that inflaton self-interactions will also be fixed by the symmetries. For (p)reheating, this feature implies that if one is interested in interactions, which determine rescattering and backreaction effects, the coefficients for these terms that appear in the action will also be fixed by the same symmetries. This paper works in the context of conformal symmetry with potential $V=\lambda \phi^4$, but subsequently allows for breaking through the inclusion of a mass term. 
		
		\subsection{Coupling to the Inflaton}
		To complete the EFT one must couple the inflationary sector to an additional reheating field, which we label $\chi(t,\vec{x})$. We are interested in the production of $\chi(t,\vec{x})$ particles resulting from the oscillations of the background inflaton field $\phi_0(t)$. In unitary gauge, the production of particles by the background will result from operators $f(t) \hat{{\cal O}}_n(\chi)$. At the quadratic level, the relevant action has the form 
		\be
		S^{(2)}_{\chi} =\int d^4x\sqrt{-g}\left[-\frac{\alpha_1(t)}{2}g^{\mu\nu}\partial_\mu\chi\partial_\nu\chi \right.
		+\left.\frac{\alpha_2(t)}{2}(\partial^o\chi)^2-\frac{\alpha_3(t)}{2}\chi^2+\alpha_4(t)\chi\partial^o\chi\right]. \;\;
		\label{sigma_action}
		\ee
		Note that the broken time translation allows for a non-trivial sound speed $c_\chi^2=\alpha_1/(\alpha_1+\alpha_2)$. As shown in \cite{Ozsoy:2017mqc,Ozsoy:2015rna}, the action of equation  \eqref{sigma_action} accounts for many existing models in the literature. For example, preheating with $V \sim g^2 \phi_0^2 \chi^2$ corresponds to $\alpha_1=1,\alpha_2=\alpha_4=0$ and $\alpha_3=g^2 \phi_0(t)^2$. If we require the inflaton to remain shift symmetric throughout reheating, as one might anticipate in models of Natural Inflation, then we consider interactions of the form $(\partial_\mu \phi_0)^2 \chi^2/\Lambda^2$, where $\Lambda$ is the cutoff for the {\it background}. The EFT captures this model through the parameter choice  $\alpha_3=2\dot{\phi}_0(t)^2/\Lambda^2$.  Previous work \cite{ArmendarizPicon:2007iv} shows that preheating is not efficient in models that preserve an inflaton shift symmetry. One reason is that naively we assume that the energy of the fields can not exceed the cutoff $\Lambda$. An important caveat to this are models with axion type couplings that have been argued to be effective \cite{Cuissa:2018oiw,Adshead:2015pva,Adshead:2018doq} where reheating can be efficient. 
		But returning to the advantage of the EFT approach is that the parameters, such as
		$\alpha_3$, can be completely non-linear, and their origin is irrelevant since the background itself is not physically observable.
		This is analogous to the EFT of Inflation, where the background is not an observable so a quasi-de Sitter background is assumed a priori, and one studies the EFT of fluctuations about that background.
		
		To conclude this section we emphasize the importance of the universality classes of the EFT as our motivation to pursue an analytical and numerical investigation into the conformal model that follows. Further work is needed to motivate this approach within a UV complete theory. Furthermore, two primary concerns with a dominant $\lambda \phi^4$ term in the potential are (a) that such a potential is ruled out by CMB observations and (b) that radiative corrections would typically generate a mass term. For the former concern, we note that the potential at the end of inflation is not the same as during inflation. For the latter concern, we conjecture that such corrections could be subdominant for some models, especially with a better understanding of the UV completion. However, we do include a mass term in the analysis in a subsequent section. 
		
		
		
		\section{Reheating Model with Nonlinear Oscillations} 
		\label{sec:model}
		
		In this section we consider preheating for a particular model. Motivated by the discussion of the previous section, we take the inflaton field to have a quartic form during reheating, and consider its coupling to the reheating field to have a simple form. Specifically, during the epoch of preheating, the potential is assumed to have the form  
		\be
		V (\phi,\chi) = {1\over4} \lambda \phi^4 + {1\over2} m_\chi^2 \chi^2
		+ \sigma \phi \chi^2 \,,
		\label{potform} 
		\ee
		where $\phi$ is the inflaton field and $\chi$ is the field being populated by the preheating process. Note (once again) that the potential during the epoch of reheating can be different from the full potential and/or the potential during the slow-roll epoch when most of the exponential expansion of the universe takes place.  For example, if the potential for the inflation takes one of the forms 
		\be
		V(\phi) = {1\over2} \Lambda^4 \left( 1 + \cos[\phi/f] \right)^2 \qquad {\rm or} \qquad V(\phi) = \Lambda^4 
		\left(1 - [\phi/f]^2 \right)^4 \,, 
		\ee
		then the leading order form of the potential will have a quartic form when $\phi$ is near its minimum.\footnote{Note that one has to translate the field $\phi\to\phi-f$, in order to obtain the first term in equation [\ref{potform}]).} Note that the potential (\ref{potform}) represents the simplest model of this type.  In particular, the interaction term is linear in $\phi$ and quadratic in $\chi$, which results in a linear equation of motion of the reheating field and requires that the coupling field $\sigma$ is sufficiently small (so that the coupling term does not interfere with the oscillations of the inflaton in its quartic potential). In addition to having a different form at earlier epochs, the inflaton potential could have additional terms (e.g., $V\sim m^2\phi^2$) during reheating, as well as coupling to additional fields. These complications are considered in subsequent sections below. 
		The interaction between $\phi$ and $\chi$ may take other forms as well, and conformal interaction terms are of particular interest considering the previous discussion. We show in Appendix \ref{sec:quartic-term-excl} that such couplings don't give favorable results to leading order, so we consider the trilinear interaction throughout this exploration of the quartic inflaton potential's effects on reheating.
		
		\subsection{Solutions for the Inflaton Field}
		\label{sec:infsolution}  
		
		In this scenario, the inflaton undergoes oscillations in a quartic potential, where we assume that its potential is dominated by the first (self-interaction) term in equation (\ref{potform}). To leading order, the resulting equation of motion has the form
		\be
		{\ddot \phi} + \lambda \phi^3 = 0 \,. 
		\ee
		This equation will have oscillatory solutions [such that $\phi(t+T)=\phi(t)$] in the absence of cosmic expansion, coupling to the $\chi$ field, and other complications. Since the equation is nonlinear, the amplitude of oscillations plays a nontrivial role. Let the oscillation have amplitude $\amp$, so that $\phi(0)=\amp$ at the start of each oscillatory cycle. We can define a reduced function $\phire$ such that   
		\be
		\phi(t) = \amp \phire(t) \qquad {\rm and} \qquad
		\phire(0)=1\,, 
		\ee
		so that $t=0$ corresponds to the beginning of a cycle.  We thus obtain the equation of motion 
		\be
		\amp {\ddot \phire} + \lambda \amp^3 \phire^3 = 0 \,. 
		\ee
		If we then define a new time variable,
		\be
		\tred = \sqrt{\lambda} \,\amp\, t \,. 
		\ee
		the equation of motion for the reduced field takes the seemingly simple form 
		\be
		{\ddot \phire} + \phire^3 = 0 \,.
		\label{reducephi} 
		\ee
		This derivation implicitly assumes that the expansion of the universe is sufficiently slow, so that the amplitude $\amp$ varies much more slowly than the field oscillates. Specifically, these solutions apply in the limit where $\amp/{\dot \amp}\gg T$. 
		
		As shown in the Appendix, the solution to equation (\ref{reducephi}) can be written exactly in terms of elliptical integrals. Since such functions are cumbersome, it is also useful to write the leading order solution in the form 
		\be
		\phire = \phire(\omega t) = \cos(\omega t)
		\left[ 1 - \epsilon \sin^2(\omega t)\right]\,,
		\label{solution} 
		\ee
		where the frequency $\omega$ is given (exactly) by equation
		(\ref{omegaright}). The corresponding period is thus $T=2\pi/\omega$. For comparison, recall that the solution for linear oscillations in a quadratic ($V\sim m^2\phi^2$) potential has the form $\phi(t)\sim\cos(mt)$. The dimensionless parameter $\epsilon$ can be found (approximated) in several ways (see Appendix \ref{sec:phisolution}). The optimum value $\epsilon\approx0.1704$ leads to an approximation for the solution $\phi(t)$ that has a relative error of about $0.36\%$. Since the frequency $\omega$ (equivalently the period $T$) is exact, the field returns to its correct value at the end of each cycle, and the error does not accumulate. As a result, equation (\ref{solution}), with known values of the parameters $(\omega,\epsilon)$, provides a sufficiently accurate solution for the nonlinear oscillations of the inflaton field. 
		
		The fluctuations of the inflaton field will experience the equation of motion
		\begin{align}
			\delta \ddot \phi_k + \left[k_\phi^2+m_\phi^2 + 3\lambda \phi^2+2\sigma\chi\phi\right]\delta \phi_k=0\ ,
		\end{align}
		eventually fragmenting the inflaton and breaking the mean-field approximation when $\langle\delta\phi^2\rangle\sim\bar\Phi^2$.
		The growth of the inflaton fluctuations can be treated similarly to the growth of the fluctuations in the reheat field.
		However, we do not consider such higher-order effects in this paper and the numeric results show that this growth is subdominant to the growth of the reheat field.

		\subsection{Reheating Field} 
		
		Following standard practice \cite{Lozanov2019}, we consider reheating into a mode of the $\chi$ field with wavenumber $k$. In the limit where the expansion of the universe can be ignored, the equation of motion for the $\chi$ field takes the form  
		\be
		{\ddot \chi}_k + \left[k^2 + m_\chi^2 + 2 \sigma \amp
		\phire(\omega t) \right] \chi_k = 0 \,,
		\label{hillkmode} 
		\ee
		where $\phire$ is the solution for oscillations of the inflaton field as given by equation (\ref{solution}).  We can change variables to make the equation appear in standard form using the definitions  
		\be
		\tau = {\omega t \over 2} \,, \qquad A_k = {4\over \omega^2}
		(k^2 + m_\chi^2)\,,
		\qquad {\rm and} \qquad q = 4 {\sigma \amp \over \omega^2} \,.
		\label{rescale} 
		\ee
		The resulting equation thus becomes
		\be
		{d^2 \chi_k \over d\tau^2} + \left[A_k + 2q \phire(2\tau)
		\right] \chi_k = 0 \,, 
		\label{hillscale} 
		\ee
		where $\phire$ is periodic so that the result is a form of Hill's equation \cite{MagWink1966}. We note that this mode equation can be written in various forms. For comparison, one common model for preheating arises for a quadratic potential, where the analog of reduced solution $\phire$ becomes $\cos(2\tau)$ and the above equation becomes the Mathieu equation (see the review of \cite{Lozanov2019}).
		
		For inflaton oscillations in a purely quartic potential, the effective equation of state parameter for the background universe has the form $\langle{w}\rangle\approx1/3$. During reheating the universe thus expands like a radiation dominated space-time with scale factor $a\sim t^{1/2}$ and $H\sim a^{-2}$.  If we include the expansion of the background into Hills equation, and work in terms of the canonically normalized field, 
		\be
		{\tilde \chi} = a \chi \,,
		\ee
		then the equation of motion (\ref{hillkmode}) becomes
		\be
		{d^2 {\tilde \chi}_k \over dt^2} + \left[ \frac{k^2}{a^2} + m_\chi^2 +
		2 \sigma \amp \phire (\omega t) - H^2 - {\dot H} \right]
		{\tilde\chi}_k = 0 \, .
		\ee
		Unlike the quadratic case, the $H^2$ and ${\dot H}$ terms do not strictly cancel. However, both terms are proportional to $a^{-4}$ and quickly redshift away compared to the other contributions. As a result, we drop these terms.  
		
		For the case of interest, where the expanding space-time acts like a radiation dominated universe, the amplitude redshifts according to $\amp\sim a^{-1}$. Now we can relabel so that $\amp_0$ denotes the non-redshifted amplitude at the start of reheating, when we set $a=1$ by definition. After removing the tildes on $\chi$ and using the time variable $\tau$ from before, we can write the resulting equation of motion in the form 
		\be
		{d^2\chi_k \over d\tau^2} + {4 \over \omega^2}
		\left[ \frac{k^2}{a^2} + m_\chi^2 + 2\sigma\amp_0 a^{-1} \phire (2\tau) \right] \chi_k = 0 \, . 
		\ee
		We thus recover Hills equation in the form
		\be
		{d^2 \chi_k \over d\tau^2} + \left( A + 2q \cos(2\tau)
		[ 1 - \epsilon \sin^2 (2\tau) ] \right) \chi_k = 0 \,.
		\ee
		where we have used the periodic function $\phire$ defined by equation (\ref{solution}).  As the universe expands, the amplitude $\amp$ decreases and the parameters ($A,q$) in Hills equation vary with time. Here, again, we assume that the expansion of the universe is slow enough that the $\chi$ field experiences many oscillations before the parameters in the equation of motion change. Since $a=1$ at the beginning of the reheating epoch, by convention, the amplitude $\amp=\amp_0/a$, and the frequency $\omega=\omega_0/a$, so that the parameters in Hills equation are given by 
		\be
		\tau={\omega_0 t \over 2a} \,, \qquad 
		A = {4a^2 \over \omega_0^2} \left( {k^2 \over a^2} + m_\chi^2 \right) \,,
		\qquad {\rm and} \qquad
		q = {4 a \sigma\amp_0 \over \omega_0^2} \,. 
		\label{paramva}
		\ee 
		In order for this scenario to remain consistent, the time scales must obey the ordering $H^{-1}\gg\omega^{-1}$ or $H \ll \omega$ (see also the discussion of Section \ref{sec:fieldtheory}). 
		
		In physical terms, the expansion of the universe must be slow compared to the oscillations of the inflaton field, so that $\phi$ experiences a large number of oscillations during the preheating epoch. The frequency of the $\phi$ oscillations is given by $\omega={\widetilde\omega}$ $\sqrt{\lambda}\,\amp$, where $\omega$ is a constant determined in Appendix \ref{sec:phisolution}, and the Hubble parameter is given by $H\sim\sqrt{8\pi\lambda/3}\,\amp^2/\mpl$. Ignoring factors of order unity, consistent solutions thus require the ordering
		\be
		\amp \ll \mpl \,.
		\label{order} 
		\ee
		The displacement of the $\phi$ field is expected to be of order the Planck scale $\mpl$ at the {\it beginning} of inflation (the beginning of slow-roll epoch), and smaller at the end of inflation (the end of the slow-roll epoch, or the beginning of the reheating epoch). As a result, the ordering of equation (\ref{order}) is expected to hold during reheating. Moreover, the amplitude $\amp$ decreases with time, so that the fidelity of the approximation increases as reheating takes place.  
		
		We can make an estimate of the total number of oscillations that the $\chi$ field experiences. This number of cycles is given by the integral 
		\be
		N_{\rm osc} = \int \omega dt = \int_1^a \omega {da\over{a}H} =
		{{\widetilde\omega} \sqrt{\lambda} \over \sqrt{8\pi\lambda/3}} 
		\int_1^a  {\mpl da \over a \amp} =
		{{\widetilde\omega}\sqrt{3}\over\sqrt{8\pi}} {\mpl\over\amp_0}
		(a-1) \,,
		\label{oscnumber} 
		\ee
		where the range of integration extends from the start of reheating(end of the slow roll phase) to the end of reheating (when parametric resonance can no longer operate). The subscript on the amplitude indicates its value at the beginning of reheating. We can estimate this value by evaluating the slow-roll conditions using the quartic potential of equation (\ref{potform}). Note that even if the inflaton potential has a different form during most of the slow-roll epoch (as expected), it must transition to the form of equation (\ref{potform}) for the reheating epoch (in the scenario considered here), so that the slow-roll conditions must be violated. This calculation yields $\amp_0\approx\mpl/\sqrt{2\pi}$. With this specification, the dimensionless coefficient in equation (\ref{oscnumber}) is of order unity and the number of oscillations is approximately given by $N_{\rm osc}\sim a$. The reheating field $\chi$ experiences the required large number of oscillations if and only if the scale factor of the universe increases by a large factor during the epoch. 
		
		\section{Parametric Resonance}
		\label{sec:parametric}
		
		One example of successful preheating after inflation occurs when the reheating field ($\chi$ in the present context) experiences parametric resonance. For the case of quartic oscillations of the inflaton field $\phi$, the previous section derives the corresponding Hills equation for the reheating field $\chi$. We thus need to determine the stability diagram to determine the conditions required for parametric resonance. We start with a brief review in order to define notation. Consider the general form of Hill's equation written in the form 
		\be
		{\ddot u} + \Omega^2 u = 0 \,,
		\ee
		where $\Omega^2$ is a periodic function with period $T$, so that $\Omega^2(t+T) = \Omega^2(t)$.  On the interval $[0,T]$, the equation has two linearly independent solutions. We choose these two solutions to be the principal solutions, which are defined such that  
		\be
		u_1(0) = 1 \qquad {\rm and} \qquad {\dot u}_1 (0) = 0
		\ee
		and 
		\be
		u_2(0) = 0 \qquad {\rm and} \qquad {\dot u}_2 (0) = 1\,.
		\ee
		Any solution can be written as a linear combination  of the two principal solutions. We conceptually break up the time evolution into intervals of length $T$. For a given interval, labeled here by the index $n$, the solution has the form 
		\be
		u_n(t) = \alpha_n u_1(t) + \beta_n u_2(t) \,. 
		\ee
		Similarly, for the next interval, labeled as $n+1$, we have
		\be
		u_{n+1}(t) = \alpha_{n+1} u_1(t) + \beta_{n+1} u_2(t) \,. 
		\ee
		If we enforce continuity of both the function and its derivative at the interface between successive cycles, we obtain two matching conditions:
		\be
		\alpha_n u_1(T) + \beta_n u_2(T) =
		\alpha_{n+1} u_1(0) + \beta_{n+1} u_2(0) = \alpha_{n+1} 
		\ee
		and 
		\be
		\alpha_n {\dot u}_1(T) + \beta_n {\dot u}_2(T) =
		\alpha_{n+1} {\dot u}_1(0) + \beta_{n+1} {\dot u}_2(0) = \beta_{n+1} 
		\ee
		These two results thus take the form
		\be
		\begin{bmatrix}
			\alpha_{n+1}  & \beta_{n+1} 
		\end{bmatrix} =
		\begin{bmatrix}
			u_1(T)  & u_2(T) \\
			{\dot u}_1(T) & {\dot u}_2(T) 
		\end{bmatrix}
		\begin{bmatrix}
			\alpha_n \\
			\beta_n
		\end{bmatrix} 
		\equiv \mathbb{M}_n 
		\begin{bmatrix}
			\alpha_n \\
			\beta_n
		\end{bmatrix} \,,
		\ee
		where the final equality defines the transfer matrix $\mathbb{M}_n$. Standard arguments \cite{MagWink1966} show that the determinant of the transfer matrix must be unity, so that the number of independent matrix elements is reduced from four to three. In addition, {\it for the case where the equation is symmetric with respect to the midpoint of the time interval}, we have the additional condition $u_1(T)={\dot u}_2(T)$.\footnote{Note that much of the literature on preheating does not take into account the symmetry condition $u_1(T)={\dot u}_2(T)$. This condition provides a useful consistency constraint on the numerical evaluation of the matrix elements, and can be used to simplify expression for the eigenvalues. }  As a result, only two of the matrix elements are independent. Here we define 
		\be
		h_n \equiv u_1(T) \qquad {\rm and} \qquad
		g_n \equiv {\dot u}_1(T) \,,
		\label{hgdefine} 
		\ee
		so that the transfer matrix has the form
		\be
		\mathbb{M}_n =
		\begin{bmatrix}
			h_n & (h_n^2-1)/g_n \\
			g_n & h_n 
		\end{bmatrix} \,. 
		\label{transfer} 
		\ee
		
		If all of the cycles are the same, then the matrix elements will be the same for all cycles, and the eigenvalues are given by 
		\be
		\lambda = h \pm \sqrt{h^2 -1}\,.
		\ee 
		After $N$ cycles, the solution to the differential equation is then given by the form 
		\be
		\begin{bmatrix}
			\alpha_N \\
			\beta_N 
		\end{bmatrix}
		= \mathbb{M}^N 
		\begin{bmatrix}
			\alpha_0 \\ 
			\beta_0 \,.
		\end{bmatrix} 
		\label{ncycle} 
		\ee
		We can rewrite the vector $[\alpha_0,\beta_0]$ in the form
		\be
		\begin{bmatrix}
			\alpha_0 \\
			\beta_0 
		\end{bmatrix}
		= A {\vec V}_1 + B {\vec V}_2 \,,
		\ee
		where the ${\vec V}_i$ are the eigenvectors of the matrix $\mathbb{M}$, corresponding to eigenvalues $\lambda_i$. Without loss of generality, we can take the largest eigenvalue to be $\lambda_1$. After $N$ cycles, the matrix/vector form of the solution becomes 
		\be
		\begin{bmatrix}
			\alpha_N \\ 
			\beta_N 
		\end{bmatrix} \approx
		\lambda_1^N A {\vec V}_1 \,.
		\ee
		If we take the limit $N\to\infty$, the growth rate for the solution is found to be 
		\be
		\gamma = {1 \over T} \log\lambda_1 \,. 
		\ee
		
		\begin{figure}[t]
			\includegraphics[scale=0.50]{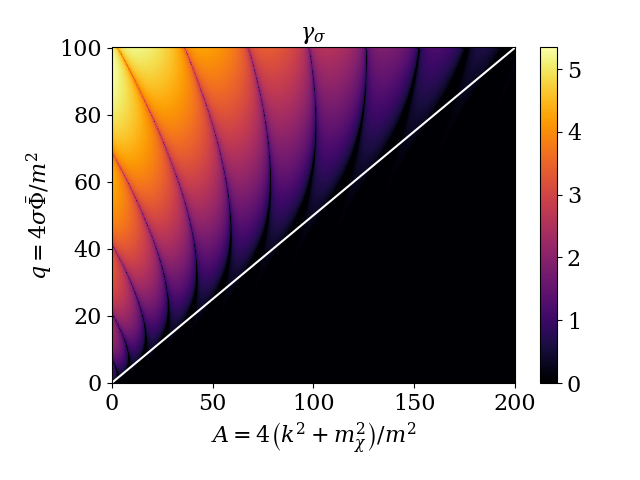}
			\caption{Stability chart in $(A,q)$ parameter space for the Mathieu equation, i.e., Hill's equation resulting from trilinear reheating and no inflaton quartic potential ($\lambda=0$ and $\epsilon=0$). The color scale shows the magnitude of the growth rate (the Floquet exponent). The dark regions correspond to near-zero growth rate and hence stability. The white $A=2q$ line separates the mostly-stable region below it from the unstable region and stability bands above. } 
			\label{fig:quartplane} 
		\end{figure}
		
		The growth rates shown in Figures \ref{fig:quartplane} and \ref{fig:quartic-trilinear-stability}, for quadratic and quartic potentials respectively, were calculated using this method.
		The similarity in the stability maps is evident, and we re-emphasize that the major difference between the models is the values that $A$ and $q$ take as the Universe expands and the resulting difference in the amount of preheating achieved: in the quartic case, the growth rate increases indefinitely while in the quadratic case the growth rate always tends toward zero.
		
		\begin{figure}
			\centering
			\includegraphics[width=0.49\textwidth]{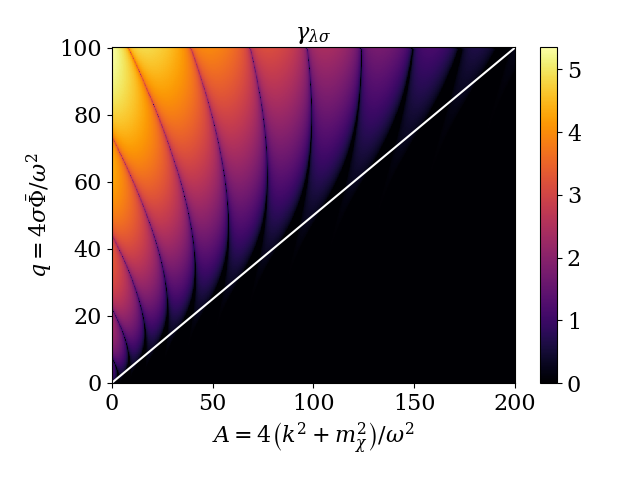}
			\includegraphics[width=0.49\textwidth]{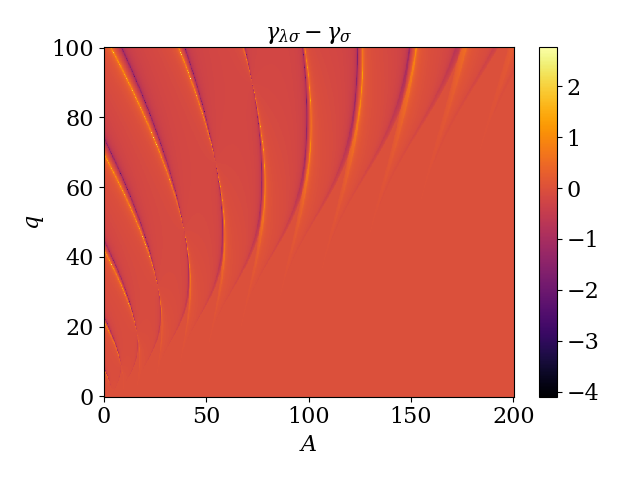}
			\caption{Stability chart in $(A,q)$ parameter space for the Hill's equation resulting from trilinear reheating with a quartic inflaton potential, $\epsilon=0.17$ (left) and the difference from the stability chart resulting from a purely massive potential, $\epsilon=0$ (right) for comparison. The white $A = 2q$ line separates the
				mostly-stable region below it from the unstable region and stability bands above.
			}
			\label{fig:quartic-trilinear-stability}
		\end{figure}
		
		For the Hills equation (\ref{hillkmode}) that arises for quartic reheating, the instability diagram is shown in Figure \ref{fig:quartic-trilinear-stability}. The color scales denotes the values of the growth rate $\gamma$ over the parameter space $(A,q)$. The diagram shows the classic pattern with bands of stability (dark in the figure) and bands of instability (light colors). As the frequency parameter $A$ increases, the Hills equation tends to become more stable. As the forcing parameter $q$ grows larger, the Hills equation becomes more unstable, although narrow bands of stability remain. The resulting instability diagram shown in Figure \ref{fig:quartic-trilinear-stability} is similar to that found for the Mathieu equation, which is often used as a model for parametric resonance in preheating \cite{Lozanov2019} and is shown in Figure \ref{fig:quartplane}. 
		
		The instability diagram shows several important features that are common to Hills equations. In addition to the classic band structure, the growth rate for instability increases toward the upper left part of the $(A,q)$ plane, and decreases toward the lower right. Although the boundary between stable and unstable regions in highly irregular, the locus $q\sim{A}/2$ provides an approximate boundary. In the limit of large $q$ (specially $q\gg{A}$), the growth rate increases with the approximate dependence $\gamma \sim \sqrt{q}$. Superimposed upon this general increase in the growth rate $\gamma$ with $q$ are bands of stability with ever-shrinking width. Significantly, one can show \cite{Weinstein} that the widths of the stable bands decrease exponentially with increasing forcing parameter $q$ (at constant $A$). 
		
		The most important difference between preheating with quartic potentials, and those considered previously with quadratic potentials, is the trajectory taken through the $(A,q)$ plane, i.e., the instability diagram. The parameters $(A,q)$ evolve as the universe expands. For the quartic case, the evolution of the the parameters is specified by equation (\ref{paramva}). In the limit of large wavenumber $k^2\gg m_\chi^2$, the frequency parameter $A$ remains nearly constant and the forcing parameter $q$ grows linearly with the scale factor. The trajectories in the parameter space depicted by the instability diagram thus move upward toward regions of greater instability.
		In the previous (quadratic) case, the evolution is toward lower values of $q$ and hence greater stability. For sufficiently large expansion factors, however, the wavenumber term redshifts away, and the frequency parameter  $A\propto a^2$, while $q \propto a$. In this regime, the increase in $q$ acts to increase the growth rate whereas the increase in $A$ acts in the opposite sense. 
		Figure \ref{fig:s-mus-tjs} shows examples of some trajectories and the corresponding growth rates for varying wavenumbers, varying initial $A$-values, and two reheat field mass options.
		When $m_\chi=0$, the growth rate increases indefinitely, while $m_\chi>0$ suppresses the growth rate after sufficient expansion because the trajectories veer toward a more stable region of the $(A,q)$ plane.
		
		\begin{figure}[t]
			\centering
			\includegraphics[width=0.4\textwidth]{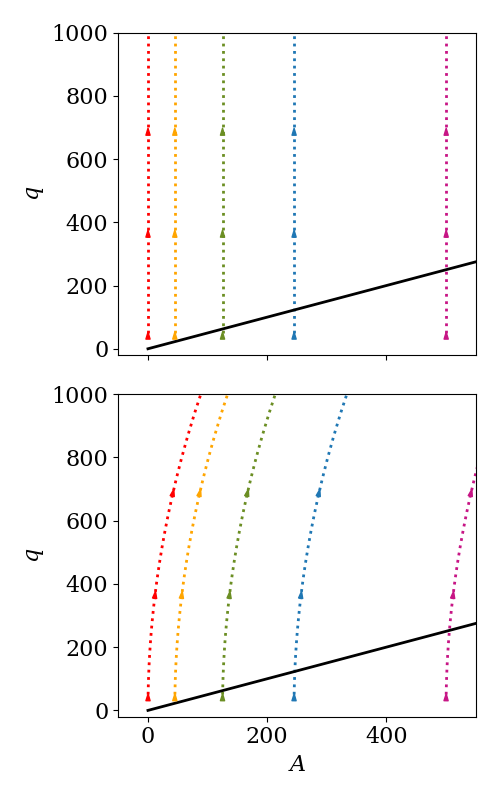}\qquad
			\includegraphics[width=0.4\textwidth]{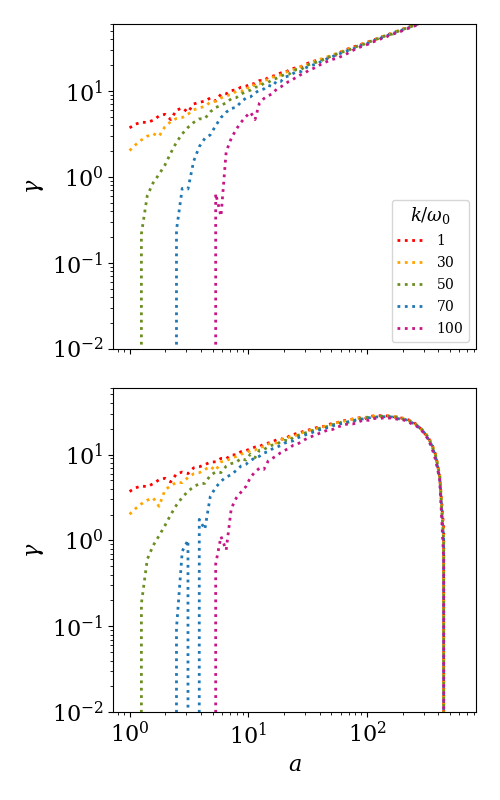}
			\caption{Left: Sample trajectories in $(A,q)$ space for $m_\chi/\omega_0=0$ (top row) and $m_\chi/\omega_0=2$ (bottom row) for trilinear quartic preheating.  The black $A=2q$ line roughly separates the unstable region above from the stable region below.  Right: corresponding growth rates $\gamma$ per oscillation time along trajectories as a function of scale factor $a$. We can see that a massless reheat field grows indefinitely without nonlinear effects, while a massive reheat field eventually stops growing on its own when the trajectory reaches a more stable area. Colors correspond to comoving wavenumbers $k/\omega_0$ chosen to result in the various $A_0$ shown, and the initial inflaton amplitude $\sqrt{\lambda}\Phi_0/\omega_0$ is chosen to result in $q_0\approx 50$.
			}
			\label{fig:s-mus-tjs}
		\end{figure}

		\section{Inclusion of a Mass Term in the Inflaton Potential}
		\label{sec:massterm} 
		
		In this section we generalize the treatment to include the case where the inflaton potential has a mass term in addition to its (dominant) quartic term.  Specifically, let the inflaton oscillations during preheating be driven by a potential of the form 
		\be
		V = {1\over2} m^2 \phi^2 + {1\over4} \lambda \phi^4 + \dots\,,
		\label{masspot}
		\ee
		so that the equation of motion becomes 
		\be
		{\ddot \phi} + m^2 \phi + \lambda \phi^3 = 0 \,. 
		\ee
		As before, we define a reduced field according to $\phi$ = $\amp \phire$, so that the equation of motion for $\phire$ has the form 
		\be
		{\ddot \phire} + \mu^2 \phire + \phire^3 = 0 \,,  
		\ee
		where we have defined dimensionless quantities 
		\be
		t_R = \sqrt{\lambda} \amp t \qquad {\rm and} \qquad
		\mu = {m \over \sqrt{\lambda} \amp} \,. 
		\ee
		To leading order, the solution has the same from as before, namely
		\be
		\phire  = \cos(\omega t) \left[ 1 - \epsilon \sin^2(\omega t) \right] \,. 
		\ee
		Again working to leading order, the parameter $\epsilon$ is given by
		\be
		\epsilon = {1 \over 9 + 8 \mu^2} =
		{1 \over 9 + 8 m^2/\lambda \amp^2} \,. 
		\ee
		The frequency $\omega$ is determined by the oscillation period, which is given by the integral expression 
		\be
		P = 4 t (\theta=\pi/2) =
		(1+\mu^2)^{-1/2} \int_0^{\pi/2} {d\theta \over 
			\left[1 - k^2 \sin^2\theta \right]^{1/2}} =
		4 \sqrt{2} k K(k) \,, 
		\ee
		where $K(k)$ is the complete elliptical integral of the first kind, and where the parameter $k$ is defined through
		\be
		k^2 \equiv {1 \over 2(1+\mu^2)}\,.
		\ee
		The frequency $\omega$ is thus given by
		\be
		\omega = {2\pi\over P} = {2 \pi \over \sqrt{2} k K(k)} 
		\ee
		The elliptical integral $K(k)$ can be expressed as
		\be
		K(k) = {\pi \over 2 \, {\rm agm} (1, \sqrt{1-k^2})} \,,
		\ee
		where agm$(x,y)$ is the arithmetic-geometric mean. To leading order we can write $K(k)$ in the form 
		\be
		K(k) \approx {\pi \over2} (1-k^2)^{-1/4} \approx
		{\pi \over2} \left( 1 + {k^2\over4} \right) \,,
		\ee
		so that the dimensionless frequency is given by
		\be
		\omega = {2\pi\over P} = {1 \over k \sqrt{2} (1+k^2/4)} \,.
		\ee
		If we convert from dimensionless form back into physical units, the frequency becomes
		\be
		\omega = {\sqrt{\lambda}\amp \over k \sqrt{2} (1+k^2/4)} =
		\sqrt{\lambda}\amp (1+\mu^2)^{1/2} {4 \over 4 + 1/(2+2\mu^2)}
		= \sqrt{\lambda}\amp {8(1+\mu^2)^{3/2} \over 9 + 8\mu^2} \,.
		\ee
		Now we can put back in the dependence of $\mu$ on the amplitude $\amp$ to obtain
		\be
		\omega = \sqrt{\lambda}\amp
		{(1+m^2/\lambda\amp^2)^{3/2} \over 9/8 + m^2/\lambda\amp^2} =
		{(m^2 + \lambda\amp^2)^{3/2} \over m^2 + (9/8) \lambda\amp^2} \,.
		\label{eqn:omega}
		\ee
		With these specifications, the $q$ parameter is given by
		\be
		q = 4 \sigma {\amp\over\omega^2} = 4\sigma\amp
		{(m^2+(9/8) \lambda\amp^2)^2 \over (m^2 + \lambda\amp^2)^3} \,,
		\ee
		and the $A$ parameter has the form 
		\be
		A = 4 \left({k^2 \over a^2} + m_\chi^2 \right) 
		{(m^2+(9/8) \lambda\amp^2)^2 \over (m^2 + \lambda\amp^2)^3} \,. 
		\ee
		The amplitude $\amp$ will redshift as the universe expands.  If the quartic part of the potential dominates, as expected at the start of preheating, then $\amp\sim1/a$. Later on, if the mass term (quadratic part of the potential) dominates, then $\amp\sim1/a^{3/2}$. The dependence of $\amp$ on $a$ depends on the equation of state parameter $\langle w\rangle$, which is given by  
		\be 
		\langle{w}\rangle = { \langle \phi (\partial{V}/\partial\phi) \rangle/2
			- \langle{V}\rangle \over
			\langle \phi (\partial{V}/\partial\phi) \rangle/2 + \langle{V}\rangle } \,.
		\ee
		For the form of the potential used here we get
		\be 
		\langle{w}\rangle = {\lambda \langle\phi^4\rangle \over
			4m^2 \langle\phi^2\rangle + 3\lambda \langle\phi^4\rangle} 
		\ee
		The expectation values of the fields can be evaluated to find
		\be
		\langle\phi^2\rangle = {\amp^2 \over 16} \left[ \epsilon^2
		- 4 \epsilon + 8 \right] \equiv f_2 (\epsilon) \amp^2 \,,
		\ee
		and
		\be
		\langle\phi^4\rangle = {\amp^4 \over 64} \left[
		{7\over16} \epsilon^4  - 3 \epsilon^3 + 9 \epsilon^2
		- 16 \epsilon + 24 \right] \equiv f_4(\epsilon) \amp^4\,,
		\ee
		where the final equalities define slowly vary functions $f_1$ and $f_2$. 
		
		\begin{figure}[t]
			\centering
			\includegraphics[width=0.8\textwidth]{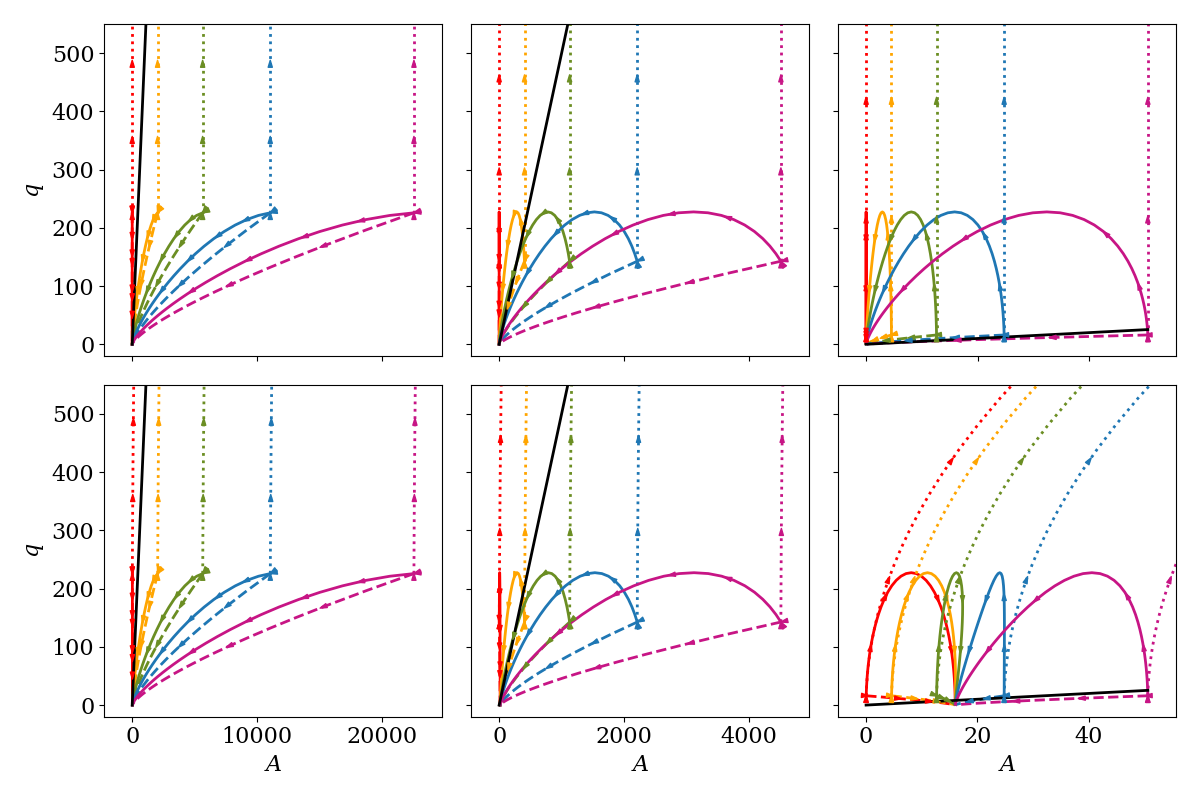}
			\caption{Trajectory examples in $(A,q)$ space for $m_\chi/\omega_0=0$ (top row) and $m_\chi/\omega_0=2$ (bottom row) for trilinear preheating.
				The black $A=2q$ line roughly separates the unstable region above from the stable region below.
				The solid lines correspond to the mixed case $m\neq0,\ \lambda\neq0$, the dashed lines to purely quartic $m=0,\ \lambda\neq0$, and the dotted lines to purely massive $m=0,\ \lambda\neq0$.
				The three columns correspond from left to right to $ \lambda\Phi_0^2/m^2=1, 10, 1000$ in the mixed potential, and the parameters for the massive and quartic cases are picked to match $(A_0,q_0)$ of the mixed cases.
				The colors correspond to different comoving wavenumbers $k/\omega_0$, listed in Figure \ref{fig:s-mus}.
			}
			\label{fig:s-tjs}
		\end{figure}
		\begin{figure}[t]
			\centering
			\includegraphics[width=0.8\textwidth]{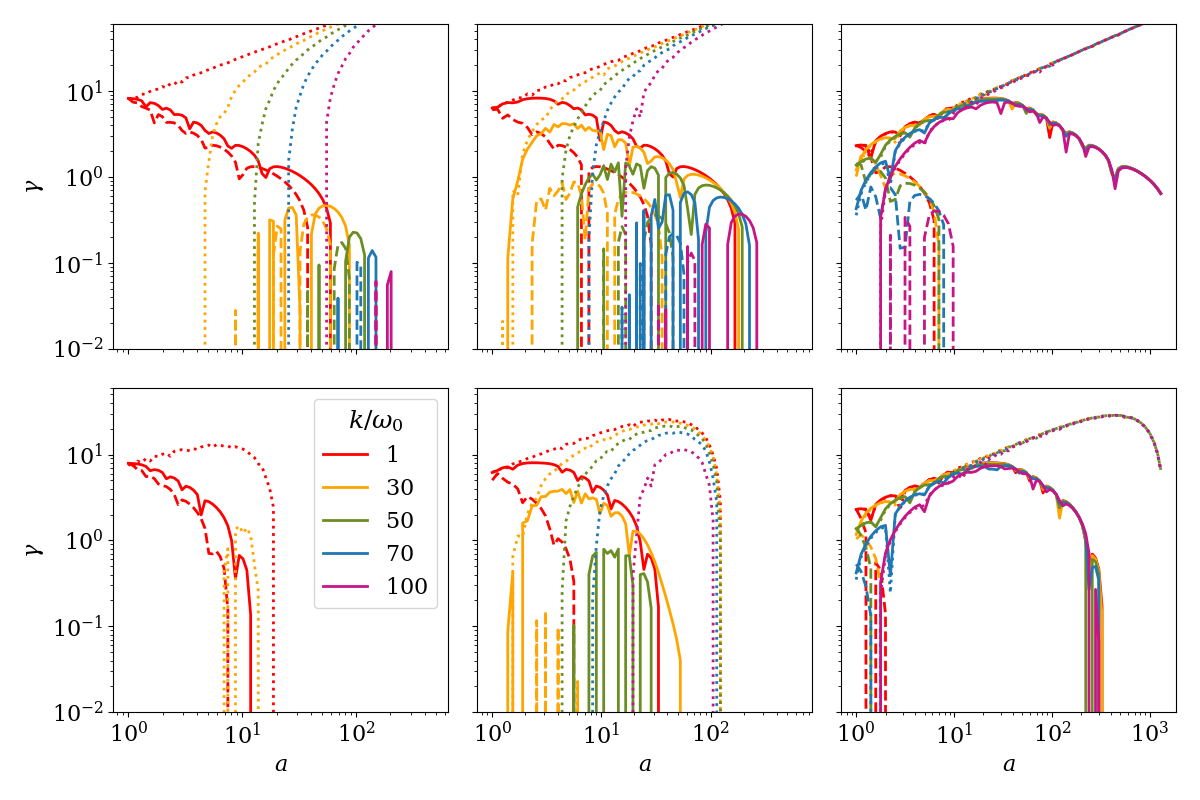}
			\caption{Growth rates $\gamma$ per oscillation time along trajectories as a function of scale factor $a$ corresponding to the trajectories in Figure \ref{fig:s-tjs}.
				The solid lines correspond to the mixed case $m\neq0,\ \lambda\neq0$, the dashed lines to purely quartic $m=0,\ \lambda\neq0$, and the dotted lines to purely massive $m=0,\ \lambda\neq0$.
				The three columns correspond from left to right to $ \lambda\Phi_0^2/m^2=1, 10, 1000$ in the mixed potential, and the parameters for the massive and quartic cases are picked to match $(A_0,q_0)$ of the mixed cases.
				The colors correspond to different wavenumbers $k_0/\omega_0$; note that for the massive case, $\omega_0=m$, for the quartic case, $\omega_0\approx 0.85\sqrt{\lambda}\Phi_0$, and for the mixed case, $\omega$ is the function shown in equation (\ref{eqn:omega}).
				We can see that mass of either the inflaton or reheat field result in an eventual decrease in the growth rate.
			}
			\label{fig:s-mus}
		\end{figure}
		
		Alternatively we can solve the continuity equation and the Friedman equation directly and find an implicit solution for the dependence of the amplitude $\amp$ on the scale factor, i.e., 
		\be
		\amp^2 (1 + u_0 \amp^2)^{1/2} = (1 + u_0)^{1/2} a^{-3}\,,
		\ee
		where $u_0\equiv\lambda\amp_0^2 f_4/(m^2 f_2)$. In deriving this result, we are assuming that the ratio $f_4/f_2$ is constant.\footnote{Since the ratio varies from 0.73 for a quartic dominated potential up to 0.75 for a quadratic dominated potential, this assumption is valid.}
		
		At the start of the reheating epoch, the quartic term in the potential dominates (for the case considered in this paper), so that $\langle w\rangle\approx1/2$, $\amp\sim1/a$, $q\sim{a}$, and  $A\sim$ $(k^2 + m_\chi^2 a^2)$. The trajectories in the $(A,q)$ plane---the instability diagram---thus move upward. For sufficiently large wavenumbers $k$, the trajectories are nearly vertical (constant $A$), whereas for lower values of $k$ the $A$ parameter increases also. At sufficiently late times, the amplitude of the inflaton oscillations are sufficiently redshifted so that $\langle w\rangle\to0$, $\amp\sim1/a^{3/2}$, $q\sim1/a$, and $A\to 4m_\chi^2/m^2$ = {\sl constant}.
		
		Figure \ref{fig:s-tjs} shows examples of trajectories in the $(A,q)$ plane for varying wavenumbers, initial $A$-values, three options of initial inflaton amplitude, and two options of the reheat field mass. In addition, trajectories for the pure quartic case, the pure quadratic case, and the mixed case are shown.
		These illustrate the major difference between quadratic and quartic inflaton potential trajectories, which is the direction of $q$-evolution.
		Figure \ref{fig:s-mus} shows the growth rates corresponding to the trajectories in Figure \ref{fig:s-tjs}.
		In all cases, the growth rates are suppressed by any of $m>0$, $m_\chi>0$, and increasing wavenumber $k$, so they ultimately determine how much preheating can occur.
		The quartic term in the inflaton potential influences the resulting reheat field spectrum, making it less redshifted;
		the larger range of amplified reheat field wavenumbers is yet another way the quartic term results in more preheating.
		With an increasing quartic influence in the potential, that is with $\lambda\neq0$ and decreasing $m$, the indefinite increase in growth rates call for higher-order effects like backreaction, rescattering, and gravitational effects to determine when reheating ends and how much particle production is achieved.
		We did not consider these and look forward to studying these effects in the future.

		\section{Numerical Results} \label{sec:numeric}
		
		Finally, we examine our scenario in fully non-linear lattice simulations.  To do this, we employ {\sc GABE} \cite{Child_2013}, which evolves the classical equations of motion for both the inflaton
		\be
		\ddot{\phi} + 3H\dot{\phi} - \frac{\nabla^2\phi}{a^2} = - \frac{\partial V}{\partial \phi}
		\ee
		and the matter field,
		\be
		\ddot{\chi} + 3H\dot{\chi} - \frac{\nabla^2\chi}{a^2} = - \frac{\partial V}{\partial \chi}
		\ee
		on a uniformly expanding grid.  The expansion of the grid is driven by the Friedmann constraint, 
		\be
		H^2 = \frac{\langle \rho\rangle}{3 m_p^2}
		\ee
		where $\langle\rho\rangle$ is the spatial average of 
		\be
		\rho = \frac{\dot{\phi}^2}{2} + \frac{\nabla^2\phi}{2a^2} + \frac{\dot{\chi}^2}{2} + \frac{\nabla^2\chi}{2a^2} + V(\phi,\chi).
		\ee
		
		For our nonlinear analysis, we focus on three specific cases: (a) the quartic inflation model, equation (\ref{potform}), where we take $\lambda = 10^{-14}$, (b) a massive inflaton, equation (\ref{masspot}), where $m_\phi = 5\times 10^{-7} m_p$ and $\lambda=0$ as well as (c) a massive inflation with a self-coupling, $m_\phi = 5\times 10^{-7} m_p$ and $\lambda=10^{-14}$.  For each of the three cases, we'll look at the impact of changing $\sigma$.  In each case, we start the simulations at the end of inflation.  To calculate the homogeneous field values, $\phi_0$ and $\dot{\phi}_0$, at this time we numerically solve the homogeneous equations of motion along the inflationary attractor until $\ddot{a}=0$.  While the Hubble scale at the end of inflation, $H_0$, is different in all three cases, it varies only slightly--between $H_0 \approx 1.93\times 10^{-7}\,m_p$ in case (b) and $H_0 \approx 2.68 \times 10^{-7}\, m_p$ in case (c).  As such we chose to keep the same physical box size at the end of inflation, $L_0 = 2.5\times 10^{-6}\,m_p$ to ease comparison between the models.  In all cases, we work with $N^3=256^3$ grids with a timestep, $\Delta t = \Delta x/40 = L_0/N/40$.  On top of the homogeneous field values, we set the fluctuations of both fields to be consistent with Bunch-Davies statistics, $\left<\phi_k^2\right>,\left<\chi_k^2\right> = 1/\sqrt{2\omega_k}$.
		
		\begin{figure}
			\centering
			\includegraphics[width=\textwidth]{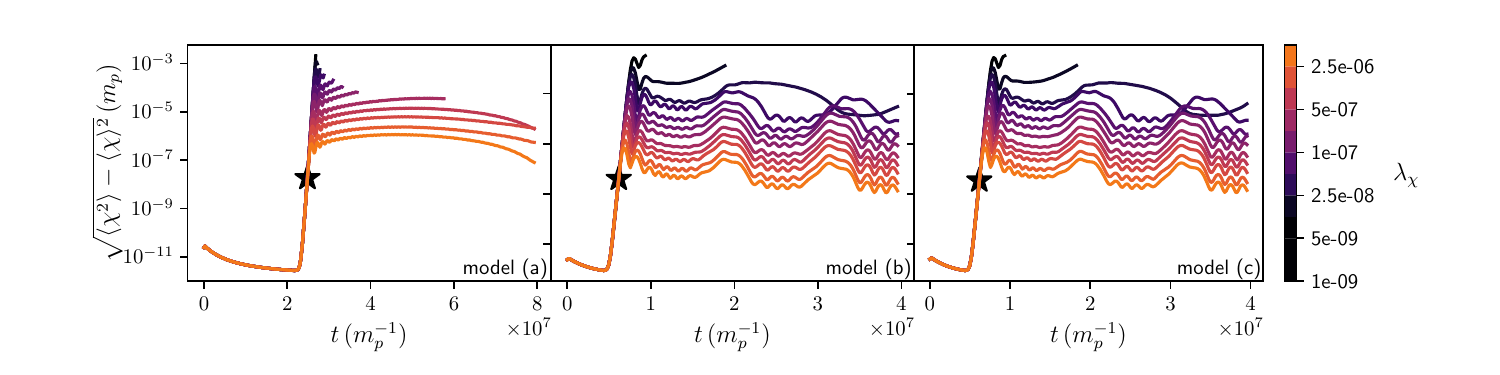}
			\caption{Variance of the matter field, $\sqrt{\langle \chi^2\rangle  - \langle \chi \rangle^2}$, as a function of time for various choices of $\lambda_\chi$.  The left panels show model (a), the middle panels show model (b) and the right panel shows model (c).  The stars indicate times in each simulation during the system is exhibiting the tachyonic instability.  All simulations set $\sigma = 2.5 \times 10^{-11}\,m_p$.}
			\label{fig:varylambda_chi}
		\end{figure}
		\begin{figure}
			\centering
			\includegraphics[width=\textwidth]{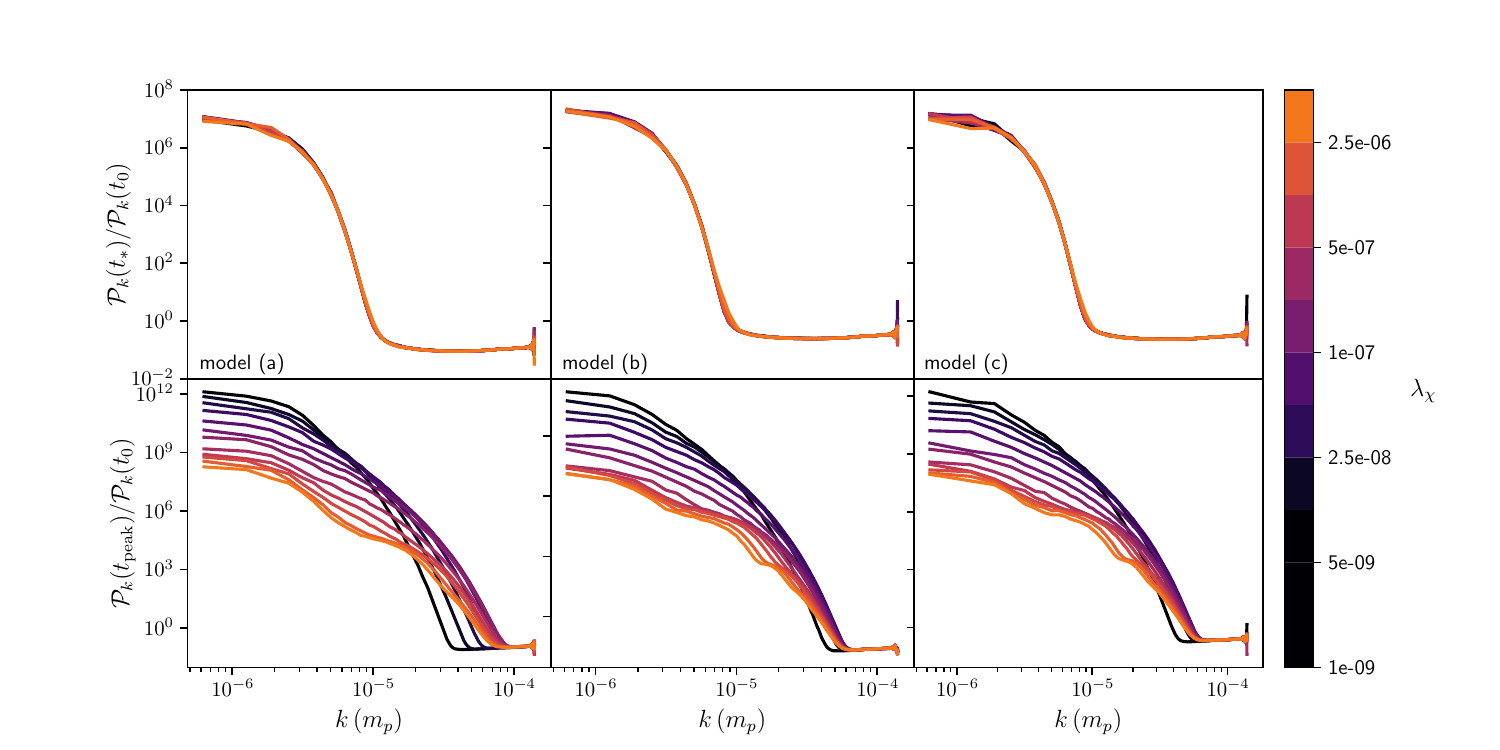}
			\caption{The top panels show the ratio of the power spectrum of $\chi$ during the tachyonic resonance phase, $t_*$, to the initial slice, $t_0$; these spectra are taken at the points labeled with stars in Figure \ref{fig:varylambda_chi}.  The bottom panels show the ratio of the power spectrum of $\chi$ at the time of the first peak of the variance of $\chi$, $t_{\rm peak}$, to the initial slice.  The left panels show model (a), the middle panels show model (b) and the right panel shows model (c). All simulations set $\sigma = 2.5\times10^{-11}\,m_p$
				\label{fig:finalspec-three}}
		\end{figure}
		
		It has long been known that studying tachyonic preheating with three-leg interactions with lattice simulations requires adding a self-coupling, $\lambda_\chi \chi^4$, to the model so that the classical potential is bounded from below \cite{Dufaux:2006ee}.  Figure \ref{fig:varylambda_chi} shows how the variance of $\chi$ evolves over the course of a set of simulations with a fixed $\sigma = 2.5\times 10^{-11}\,m_p$; as the homogeneous mode of $\phi$ changes sign for the first time, the instability in $\chi$ causes a dramatic rise in this quantity.  
		
		Importantly, during this first phase, the $\chi$ self-coupling does not have an effect on the dynamics: this can be seen by the overlapping behavior of the variance in Figure \ref{fig:varylambda_chi} as well as by studying the power spectra of $\chi$.  The top row of Figure \ref{fig:finalspec-three} shows the growth of modes by comparing the power spectrum during the tachyonic phase to the power spectrum at the beginning of the simulation.  For a wide range of choices of the self-coupling, these ratios are nearly identical.  The bottom row of Figure \ref{fig:finalspec-three} shows the effect that the self-coupling has by looking at the growth of modes until the first peak of the variance of $\chi$; that is, {\sl after} the tachyonic instability causes the $\chi$ field to enter the nonlinear regime, the self-coupling of $\chi$ has a dramatic effect on the distribution of power in the $\chi$ field.
		
		\begin{figure}
			\centering
			\includegraphics[width=\textwidth]{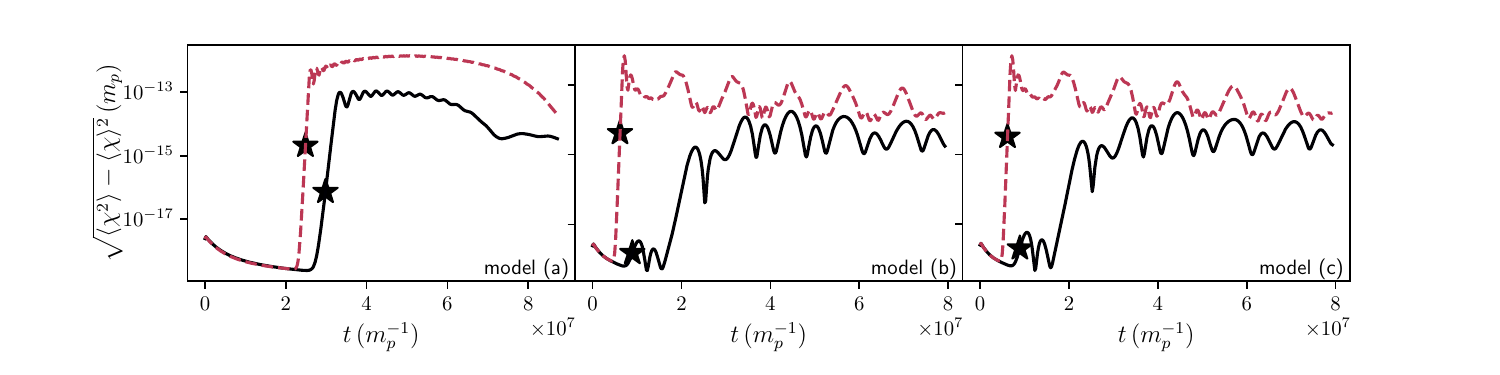}
			\caption{Variance of the matter field, $\sqrt{\langle \chi^2\rangle  - \langle \chi \rangle^2}$, as a function of time for two choices of the trilinear coupling:  $\sigma = 5\times 10^{-13}$ (black, solid) and $\sigma = 5\times 10^{-12}$ (red, dashed) .  The left panels show model (a), the middle panels show model (b) and the right panel shows model (c).  The stars on the plots show times at which we compare the power spectra in Figure \ref{fig:finalspec}.}
			\label{fig:varysigma}
		\end{figure}
		\begin{figure}
			\centering
			\includegraphics[width=.6\textwidth]{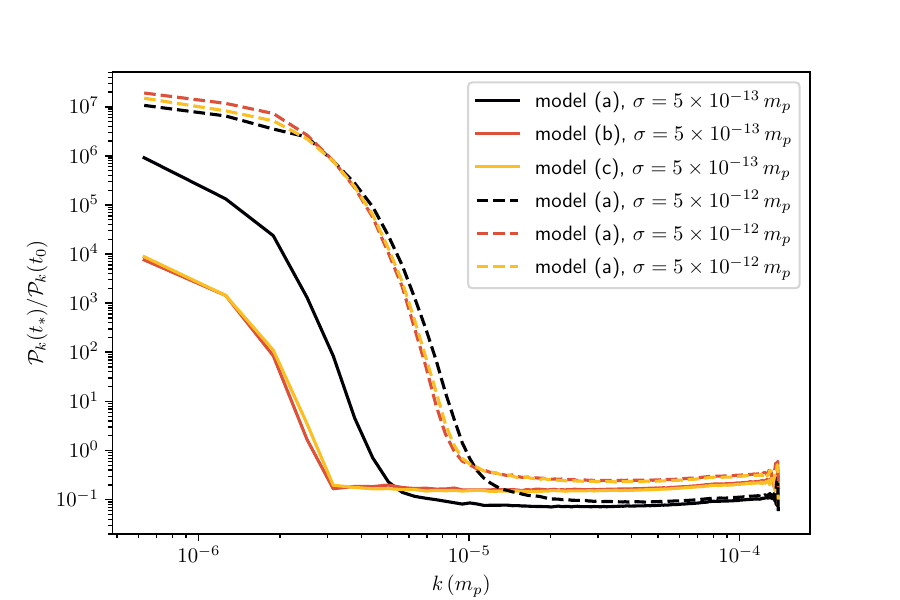}
			\caption{Ratio of the power spectrum of $\chi$ during the first resonance phase, $t_*$, to the initial slice, $t_0$; these spectra are taken at the points labeled with stars in Figure \ref{fig:varylambda_chi}.  We directly compare the spectra during this initial instability for all three models at two different values of the trilinear coupling: $\sigma = 5\times 10^{-13}$ (solid) and $\sigma = 5\times 10^{-12}$ (dashed).}
			\label{fig:finalspec}
		\end{figure}
		
		In the cases in Figures \ref{fig:varylambda_chi} and \ref{fig:finalspec-three}, the instability is very strong and we see little difference between the three cases.  To see the effect of the inflaton potential on the dynamics, we compare the fiducial $
		\sigma = 5\times 10^{-12}\,m_p$ to a lower coupling, $
		\sigma = 5\times 10^{-13}\,m_p$.  Figure \ref{fig:varysigma} shows that, for this lower choice of the coupling, $\sigma$, reheating does not always complete during the first oscillation of the field.  Additionally, we see that the instability is now model-dependent.

		We can study the initial instability for each of the three models by looking at the amplification of the power spectrum at the time when the instability occurs (labeled with stars in Figure \ref{fig:varysigma}) to the initial power spectra in each case\footnote{Note that here we directly plot the ratio of the power spectrum of the field; modes that are not amplified are expected to decrease in amplitude due to the expansion of the universe in each case.  Alternatively, we could have chosen to scale the fields to the canonically-normalized fields so that unamplified modes remain constant, as in \cite{Adshead:2015pva}.}  Fig. \ref{fig:finalspec} shows that, for this smaller trilinear coupling, we see a wider and stronger amplification in model (a) as compared to models (b) and (c) which show consistent amplification.

		\section{Conclusion}
		\label{sec:conclude} 
		
		This paper has considered reheating in inflationary scenarios where the effective potential of the inflaton has a quartic form at the end of inflation representing an approximate conformal symmetry within the EFT of reheating. 
		For this case, we find the following results: 
		
		$\bullet$ The oscillations of the inflaton field are nonlinear. The exact solutions can be described in terms of elliptical integrals, but we find an accurate approximation scheme that allows the solutions to be expressed in terms of elementary functions. 
		
		$\bullet$ The equation of motion for the coupled reheating field $\chi$ becomes a Hill's equation (instead of the more common and less general Mathieu equation) in this scenario. Over the parameter space $(A,q)$, the growth rate for parametric resonance shows the same characteristic bands of stability and instability (see Figure \ref{fig:quartic-trilinear-stability}). The shapes of the stable and unstable regions are quantitatively different than for the Mathieu equation, but qualitatively similar. 
		
		$\bullet$ The parameters  $(A,q)$ that determine parametric resonance vary with the amplitude $\amp$ of the inflaton oscillations, where $\amp$ redshifts with time (increasing scale factor). Significantly, however, the forcing parameter $q$ grows as the amplitude $\amp$ decreases, in contrast to previously considered cases, so that parametric resonance becomes more pronounced with time. 
		
		$\bullet$ This scenario for reheating can be generalized to include both quartic and quadratic terms in the effective potential for the inflaton during reheating. In this case, we find modified solutions for the nonlinear oscillations of the inflaton. The evolution of the parameters $(A,q)$ grows with decreasing amplitude $\amp$ at the start of reheating, and then subsequently decrease as the quadratic part of the inflaton potential dominates and the oscillations become linear. 
		
		
		The results of this paper show that the behavior of parametric resonance during the epoch of reheating can be more complicated than considered previously. Whereas the parameters $(A,q)$ in the Mathieu equation for quadratic potentials evolve to smaller values and hence greater stability, the corresponding parameters for quartic potentials evolve to larger values and can lead to greater instability (in addition the Mathieu equation is generalized to become Hill's equation). The trajectories through the $(A,q)$ plane are more complicated for potentials containing both quadratic and quartic components, so that the parameter values can both increase and decrease with time.
		While this paper has demonstrated these complexities, the range of allowed potentials is large, and a variety of reheating scenarios can be considered in future work.

		\appendix
		\section{Solution for Nonlinear Oscillations of the Inflaton Field}
		\label{sec:phisolution} 
		
		In this Appendix, we take up the task of solving for $\phire(t)$ in (\ref{reducephi}) which is a nonlinear equation. Although the equation of motion has a closed form solution in terms of elliptical integrals, here we develop a more tractable working approximation (see also \cite{Boyanovsky1996,Frasca2011}). We can immediately take the first integral to find 
		\be
		{1\over2} {\dot \phi}^2 = E - {1\over 4} \phi^4 \,,
		\ee
		where we now ignore the subscripts. Since we want
		the solution where $\phi(0)=1$ and ${\dot\phi}(0)=0$,
		we can specify the integration constant to find
		\be
		\sqrt{2} {\dot\phi} = \left[1 - \phi^4\right]^{1/2} \,. 
		\ee
		This equation has the formal solution of the form
		\be
		t = \sqrt{2} \int_\phi^1 {d\phi \over
			\left[1 - \phi^4 \right]^{1/2}} \,.
		\label{timeint} 
		\ee
		Since the integral in equation (\ref{timeint}) defines the dimensionless period, we can find the dimensionless frequency ${\widetilde\omega}$ through the relation 
		\be
		{\widetilde\omega} = {2\pi\over 4\sqrt{2}I} \qquad {\rm where} \qquad
		I \equiv \int_0^1 {dx \over (1-x^4)^{1/2}} = K(-1) \approx 1.311 \,,
		\label{omegaright} 
		\ee
		where $K(m)$ is the complete elliptical integral of the first kind \cite{Abstegun1972}. Keep in mind that the time variable in the solution is dimensionless, as we have scaled out a factor of $\sqrt{\lambda}$ and another factor of $\amp$. If we work in terms of the usual time variable (again denoted as $t$), then the physical frequency $\omega$ becomes 
		\be
		\omega = \sqrt{\lambda}\,\amp\,{\widetilde\omega} \approx 0.85
		\sqrt{\lambda}\,\amp \,. 
		\ee
		
		\subsection{Leading Order Solution}
		\label{sec:leadingorder} 
		
		The results thus far are exact. In order to find a working
		approximation, we expand the integral and change variables such that $\phi=\cos\theta$ to get
		\be
		t = \sqrt{2} \int_0^\theta
		{d\theta \over \left[2 - \sin^2\theta \right]^{1/2}} = 
		\int_0^\theta d\theta \left\{1 + {1\over4} \sin^2\theta + \dots
		\right\} \,.
		\label{qexpand} 
		\ee
		Now we can perform the integral to obtain
		\be
		t = {9 \over 8} \theta - {1\over16} \sin2\theta =
		{9\over8} \cos^{-1}\phi - {1\over8} \phi \sqrt{1-\phi^2} \,.
		\ee
		Note that the final term is small compared to the first. It vanishes at both $\phi=0$ and $\phi=1$, with a maximum in between, and has a coefficient nearly an order of magnitude smaller. We can thus rearrange the expression to take the form
		\be
		\phi = \cos\left[ {8\over9}t\right]
		\cos\left[{1\over9} \phi \sqrt{1-\phi^2} \right] - 
		\sin\left[ {8\over9}t\right]
		\sin\left[{1\over9} \phi \sqrt{1-\phi^2} \right] \,,
		\ee
		which becomes
		\be
		\phi \approx \cos\left[ {8\over9}t\right] -
		\sin\left[ {8\over9}t\right] {1\over9} \phi \sqrt{1-\phi^2} \approx 
		\cos\left[ {8\over9}t\right] \left\{ 1 - {1\over9} 
		\sin^2\left[ {8\over9}t\right] \right\} \,.
		\ee
		As a result, oscillations in the quartic potential can be described by solutions of the form
		\be
		\phi(t) = \amp \cos (\omega t) 
		\left[ 1 - \epsilon \sin^2 (\omega t) \right] \,. 
		\ee
		The leading order approximation given here estimates the values of $\omega = 8/9$ and $\epsilon = 1/9$. However, we can find a better approximation by using the exact value of the dimensionless frequency ${\widetilde\omega}\approx0.85$ found above in equation [\ref{omegaright}] (instead of $\omega$ = 8/9 $\approx$ 0.89). We can also find an optimum value for the parameter $\epsilon$ as shown below. 
		
		\begin{figure}
			\includegraphics[scale=0.50]{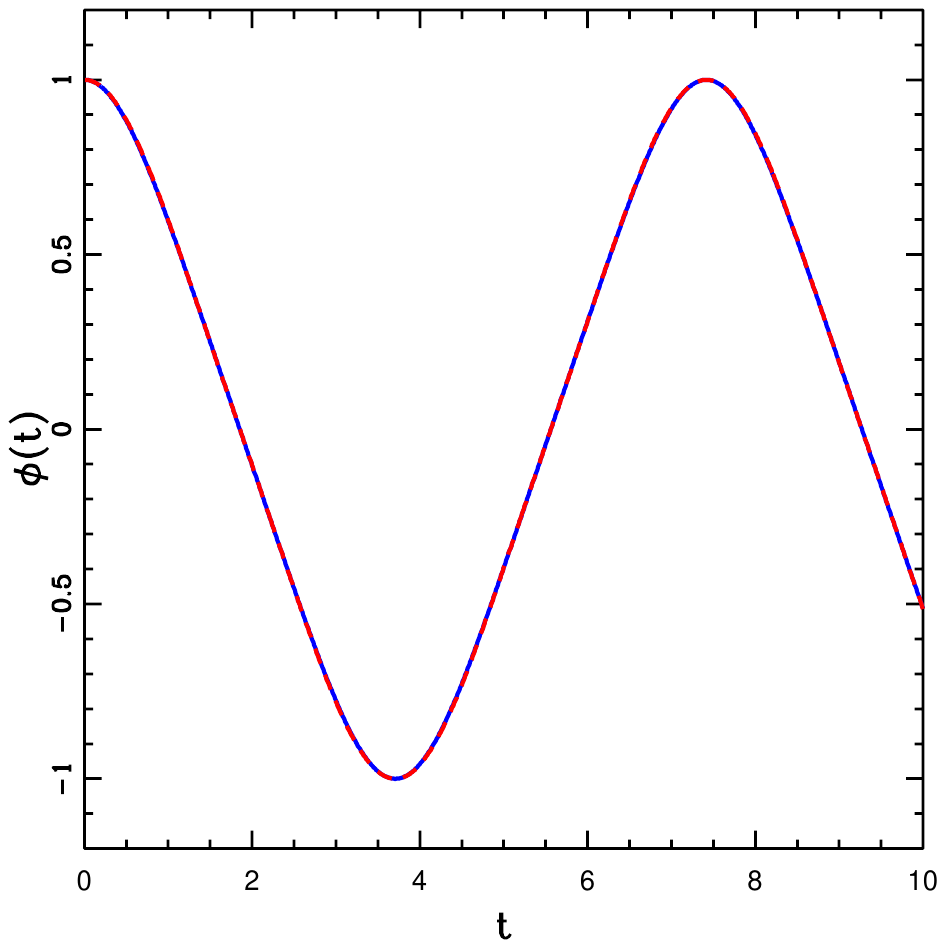}
			\vskip-1.20truein
			\caption{Solution for oscillatory wave form in quartic potential. Solid blue curve shows the exact result from numerical integration. Dashed red curve shows the approximation derived in this Appendix, where we use equation (\ref{omegaright}) to specify $\omega$ and we use the $\epsilon$ parameter that minimizes the action ($\epsilon\approx0.1704$).} 
			\label{fig:quartcomp} 
		\end{figure}
		
		\subsection{Optimization of the Correction Parameter} 
		\label{sec:opepsilon} 
		
		Here we develop a formal method to specify the optimum value of the correction parameter $\epsilon$ that characterizes the approximate solution for the field $\phi(t)$. After scaling out the amplitude, the solution has the form 
		\be
		\phi(t) = \cos(\omega{t}) \left[ 1 - \epsilon \sin^2(\omega{t})
		\right] \,,
		\label{phiform} 
		\ee
		where $\omega$ is a known quantity (in this case $\omega$ = 0.8472$\dots$). The equation of motion for $\phi$ can be derived from an action, which has the form
		\be
		S = \int_0^{2\pi} \omega{dt} \left\{ {1\over2} {\dot\phi}^2 - {1\over4}
		\phi^4 \right\} \,,
		\ee
		where we have added a factor of $\omega$ to make the result dimensionless. The usual procedure is to derive the equation of motion from the action (which results from its minimization) and then find an approximate solution. Instead, we use the form of the approximate solution and evaluate the action $S$, which provides us with a function of the parameter $\epsilon$. By minimizing the function, we can find the optimal value for $\epsilon$. We thus have to evaluate the expression 
		\be
		S = \int_0^{2\pi} d\theta \left\{ {1\over2} \omega^2 \sin^2\theta
		\left[ -1 + \epsilon - 3 \epsilon \cos^2\theta \right]^2 
		- {1\over4} \cos^4\theta 
		\left[ 1 - \epsilon \sin^2 \theta \right]^4 \right\} \,. 
		\ee
		After performing the integrals, we obtain
		\be
		S = {\pi\omega^2\over16} (8 - 4\epsilon + 5 \epsilon^2)
		- {\pi\over2048}\left[ 384 - 256\epsilon + 144 \epsilon^2
		- 48 \epsilon^3 + 7 \epsilon^4 \right] 
		\ee 
		If we take the derivative with respect to $\epsilon$, set
		the result equal to zero, and divide out a factor of 4, we obtain the condition
		\be
		64 \omega^2 (-2 + 5 \epsilon) = - 64 + 72 \epsilon -
		36 \epsilon^2 + 7 \epsilon^3 \, .
		\ee
		If we use the known value for $\omega$ (see above), and solve the resulting cubic equation, we find the optimal value $\epsilon\approx0.17036$. Note that we can also numerically solve the equation of motion for $\phi(t)$, and then find the optimal value of $\epsilon$ by fitting the result to a function of the form (\ref{phiform}). This procedure produces the same value of $\epsilon$. 
		
		Figure \ref{fig:quartcomp} illustrates the efficacy of this approximation scheme. The solid blue curve shows the  numerically determined solution for $\phi(t)$, whereas the red dashed curves shows the approximation developed here. The two curves are nearly indistinguishable, with the difference smaller than the width of the lines used for plotting.

		\section{Comparison of growth rate and expansion rate} \label{sec:gammaH}
		
		Reheating is considered efficient when $\gamma>H$, when the growth rate is greater than the expansion rate of the Universe.
		Therefore it is worth considering the evolution of the ratio $\gamma/H$;
		we find that this perspective reinforces the advantage of the quartic potential in increasing the production of $\chi$.
		
		\begin{figure}[t]
			\centering
			\includegraphics[width=0.49\linewidth]{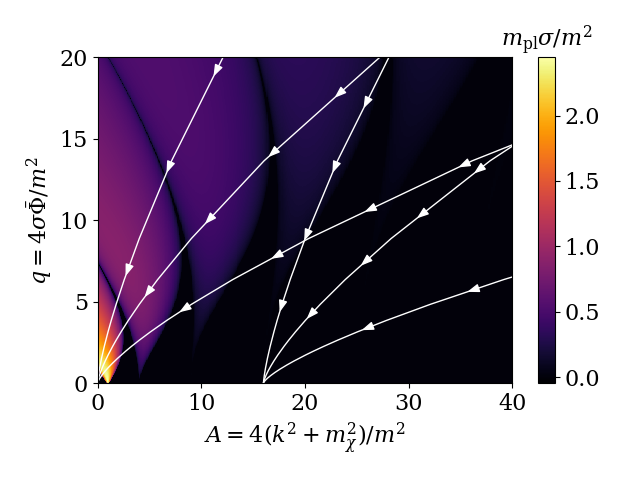}
			\includegraphics[width=0.49\textwidth]{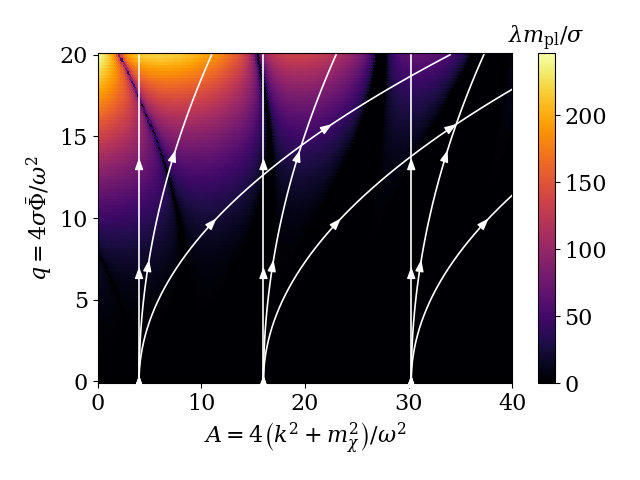}
			\caption{Map of $\gamma/H$ for $V(\phi)= \frac{1}{2} m^2\phi^2$ (left) and $V(\phi)=\frac{1}{4}\lambda \phi^4$ (right). The sample trajectories showcase the general reheating behavior, with trajectories on the left converging on narrow resonance and trajectories on the right tending toward broad resonance. These maps reinforce our conclusion that reheating with a quartic potential is much more efficient than with a massive potential, with some accessory dependence on the relative values of $m$, $\sigma$, and $\lambda$.}
			\label{fig:gammaH}
		\end{figure}
		
		Figure~\ref{fig:gammaH} shows maps of $\gamma/H$ in the $(A,q)$ plane. 
		In the case where $V(\phi)=\frac{1}{2}m^2\phi^2$, we see that $\gamma/H$ can actually increase with time, except when crossing narrow stability bands, unlike $\gamma$ alone which decreases with time.
		Even so, except for a possible burst in the bottom left corner of the map, for most of reheating we have $\gamma/H/(m_\mathrm{pl}\sigma/m^2)<1$.
		All trajectories converge on $\Phi\ll m^2/\sigma$, meaning $\gamma/H\to0$.
		In the case where $V(\phi)=\frac{1}{4}\lambda\phi^4$, trajectories can easily achieve $\gamma/H/(\lambda m_\mathrm{pl}/\sigma)\gg1$, possibly indefinitely.
		We know $\gamma\sim\sqrt{q}$ for $q\gg1$, and it's clear that $\gamma/H$ increases more quickly with time than $\gamma$ itself.
		In the case of the quartic potential, we can be more concrete since $\gamma\sim\sqrt{q}$ for large $q$, so that $\gamma/H\sim a^{5/2}$.
		In all cases, a nonzero reheat field mass $m_\chi$ suppresses growth eventually.
		
		Considering $\gamma/H$ further emphasizes the effect of the form of $V(\phi)$ on the efficiency of reheating.
		The difference can be exaggerated or tempered depending on the value of $\sigma$ relative to the other couplings $m^2$ and $\lambda$---notice the units on the colorbars in Figure \ref{fig:gammaH} and their opposite dependence on $\sigma$.

		\section{Exclusion of quartic terms} 
		\label{sec:quartic-term-excl}
		
		The most general conformal potential of two scalar fields, plus the trilinear term, is
		\begin{align}
			V(\phi,\chi) = \frac{m^2}{2}\phi^2 + \frac{\lambda}{4}\phi^4 + \frac{m_\chi^2}{2}\chi^2 + \frac{\lambda_\chi}{4}\chi^4 + \sigma\phi\chi^2 + g_1 \phi^3\chi + g_2\phi^2\chi^2 + g_3\phi\chi^3\ ,
		\end{align}
		which includes the first, second, and third quartic couplings $g_1$, $g_2$, and $g_3$, respectively.
		In this appendix we justify excluding these interactions and the quartic reheat field potential when considering only leading order effects in efficient reheating.
		
		\subsection{Quartic reheat field potential}
		
		A term $V(\chi) \supset \frac{1}{4}\lambda_\chi\chi^4$ adds a strong restorative force that suppresses the exponential increase in the reheat field's amplitude, even with modest strength $\lambda_\chi<1$.
		The trilinear equation of motion becomes, with the familiar definitions of $A$ and $q$,
		\begin{align*}
			0&=\chi_k'' + \lambda_\chi \chi_k^3+\left(\frac{4(k^2+m_\chi^2)}{\omega^2} + \frac{8\sigma\bar\Phi_0}{\omega^2} \phi_R \right) \chi_k\\
			&\equiv \chi_k'' + \left(A + \lambda_\chi\chi_k^2 + 2q\phi_R\right)\chi_k\ .
		\end{align*}
		The coefficient increases, $A\to A_\mathrm{eff}\equiv A+\lambda_\chi\chi_k^2$, which stabilizes the growth in $\chi_k$.
		The corresponding growth rates are shown in Figure \ref{fig:chi4-map-example}. 
		Therefore, we exclude significant $\lambda_\chi$ from our analyses to hasten the reheating process.
		Nevertheless, an appropriate quartic reheat potential guarantees bounding of the potential \cite{Dufaux_2006}, a feature we employed to regularize our numerical results in Section \ref{sec:numeric} with $\lambda_\chi\ll1$.

		\begin{figure}[t]
			\centering
			\includegraphics[width=0.5\textwidth]{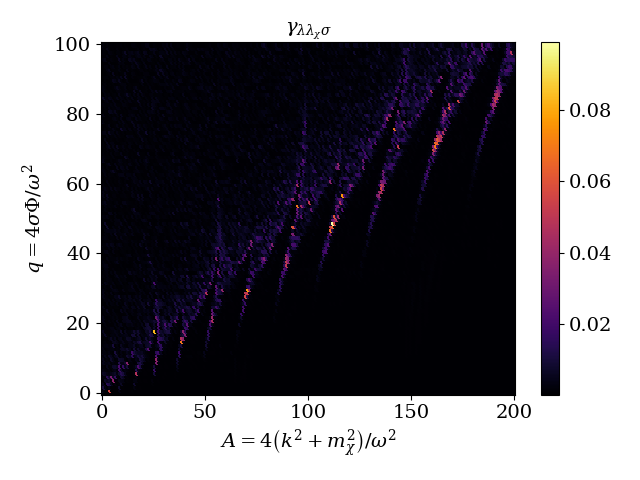}
			\caption{Stability chart in $(A,q)$ parameter space for the Hill's equation resulting from quartic reheating ($\epsilon=0.17$) through a trilinear interaction and a quartic reheat field potential with $\lambda_\chi=0.01$.}
			\label{fig:chi4-map-example}
		\end{figure}
		
		\subsection{Forced oscillator solution for first quartic coupling}
		
		The equation of motion (EOM) from the potential
		\begin{align}
			V(\phi,\chi)=\frac{m^2}{2}\phi^2 +\frac{\lambda}{4}\phi^4 + \frac{m_\chi^2}{2}\chi^2 + g_1 \phi^3\chi
		\end{align}
		is
		\begin{align}
			\ddot\chi_k + &\left(k^2+m_\chi^2\right)\chi_k =- g_1\bar\Phi^3\phi_R^3 \\
			&= -g_1\bar\Phi^3 \left( \frac{12-6\epsilon}{16} \cos (\omega t) + \frac{4+3\epsilon}{16}\cos(3\omega t) + \frac{3\epsilon}{16} \cos(5 \omega t) \right) + \mathcal{O}(\epsilon^2) \label{eqn:FO-eps}
		\end{align}
		
		The righthand side expands into a series of forcing terms, and the equation is analytically solvable.
		We can get a good idea of the solution by considering terms up to $\mathcal{O}(\epsilon)$.
		Figure \ref{fig:first-quartic} shows that the numeric solution to the full EOM visually matches the analytic particular solution to equation (\ref{eqn:FO-eps}) alone, which is
		\begin{align}
			\nonumber {\chi_k}_p &= g_1 \bar\Phi^3 \left(\frac{12-6\epsilon}{16}\ \frac{\cos (\omega t)}{k^2+m_\chi^2-\omega^2}+\frac{4+3\epsilon}{16}\ \frac{\cos (3\omega t)}{k^2+m_\chi^2-9\omega^2}+\frac{3\epsilon}{16}\ \frac{\cos (5\omega t)}{k^2+m_\chi^2-25\omega^2}\right)\\
			&\hphantom{=}+\mathcal{O}(\epsilon^2)\ \text{ for }\ k^2+m_\chi^2 \neq \omega^2, 9\omega^2, 25\omega^2\ .
		\end{align}
		This solution is only valid once again under the assumption that $\omega \gg H$ so that we can approximate $\omega$ as constant during each oscillation period.
		This solution is almost always stable.
		For instantaneous moments during reheating when we might have $(k^2+m_\chi^2)/\omega^2 = 1, 9, 25$, the growth in the amplitude of $\chi_k$ is still only linear in time.
		For example,
		\begin{align}
			\nonumber {\chi_k}_p &= g_1 \bar\Phi^3 \left(\frac{12-6\epsilon}{16}\ \frac{t\sin(\omega t)}{2\omega} +\frac{4+3\epsilon}{16}\ \frac{\cos (3\omega t)}{k^2+m_\chi^2-9\omega^2}+\frac{3\epsilon}{16}\ \frac{\cos (5\omega t)}{k^2+m_\chi^2-25\omega^2}\right) + \mathcal{O}(\epsilon^2) \\
			&\text{ for }\ k^2+m_\chi^2=\omega^2\ .
		\end{align}
		This linear growth, an example of which is shown in Figure \ref{fig:first-quartic}, is negligible compared to the exponential growth from the trilinear potential.
		\begin{figure}[t]
			\centering
			\includegraphics[width=0.49\textwidth]{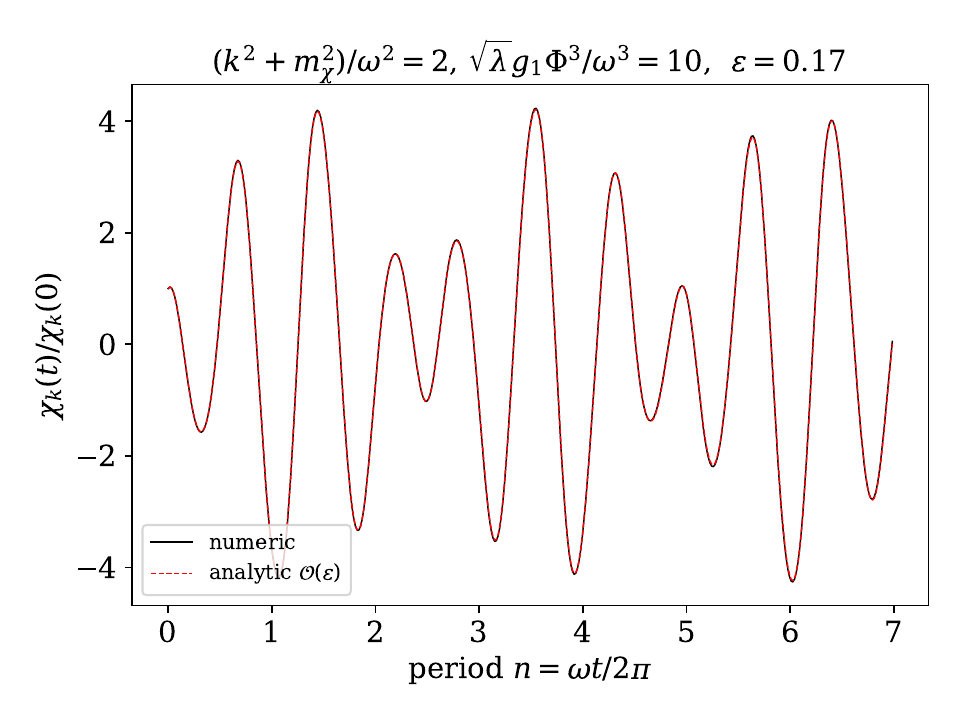}
			\includegraphics[width=0.49\textwidth]{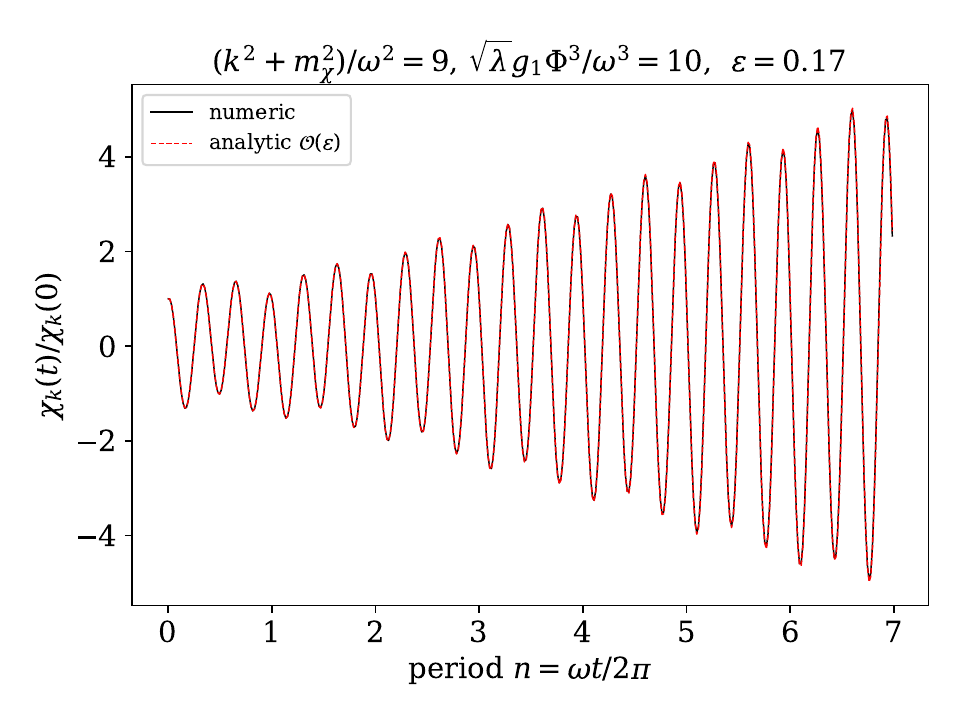}
			\caption{Example solutions for the first quartic interaction at a non-resonant (right) and resonant frequency (left) frequency using the optimal value of $\epsilon=0.17$ that enters the sourcing inflaton. At resonant frequencies, the amplitude of the reheat field grows linearly, which is nonzero but much slower growth than the exponential growth resulting from the trilinear interaction.}
			\label{fig:first-quartic}
		\end{figure}

		\subsection{Subdominant growth from second quartic coupling}

		The equation of motion from the potential
		\begin{align}
			V(\phi,\chi)=\frac{m^2}{2}\phi^2 + \frac{\lambda}{4}\phi^4 + \frac{m_\chi^2}{2}\chi^2 + g_2 \phi^2\chi^2
		\end{align}
		is
		\begin{align}
			0 &= \ddot \chi_k + \left(k^2 + m_\chi^2 + 2g_2 \bar\Phi^2\phi_R^2 \right) \chi_k \\
			0&= \chi''_k + \bigl(A+2q\cos(2\tau) [1+\epsilon\cos(2\tau)]\bigr)\chi_k + \mathcal{O}(\epsilon^2)\\
			A &= \frac{k^2+m_\chi^2+g_2(1-\epsilon)\bar\Phi^2}{\omega^2}\ ,\quad q=\frac{g_2\bar\Phi^2}{2\omega^2}
			\ ,\quad 	\omega =  \frac{(m^2+\lambda\bar\Phi^2)^{3/2}}{m^2 + \frac{9}{8}\lambda\bar\Phi^2} \ ,\quad
			\epsilon = \frac{\lambda\bar\Phi^2}{8m^2+9\lambda\bar\Phi^2}
		\end{align}
		\begin{figure}[t]
			\centering
			\includegraphics[width=0.49\textwidth]{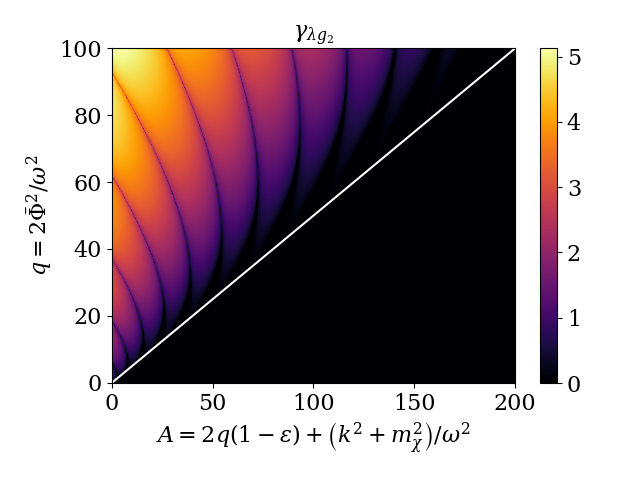}
			\includegraphics[width=0.49\textwidth]{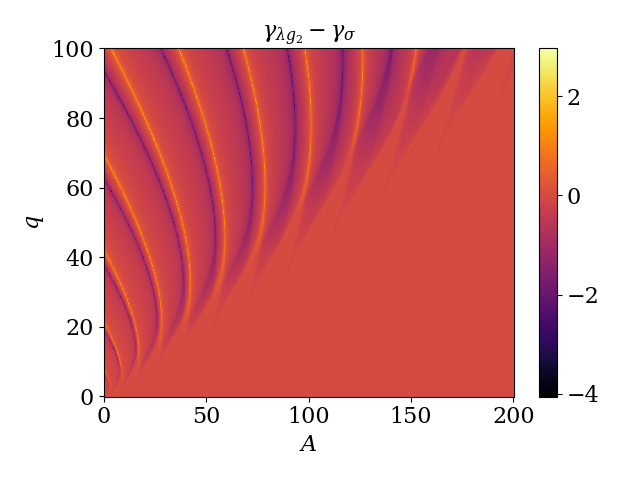}
			\caption{Stability chart in $(A,q)$ parameter space for the Hill's equation resulting from second quartic reheating with a quartic inflaton potential, $\epsilon=0.17$ (top) and the difference from the stability chart resulting from trilinear preheating with a purely massive potential, $\epsilon=0$ (bottom) for comparison. The white $A = 2q$ line separates the
				mostly-stable region below it from the unstable region and stability bands above.
			}
			\label{fig:quartic-g2-stability}
		\end{figure}
		\begin{figure}[t]
			\centering
			\includegraphics[width=0.7\textwidth]{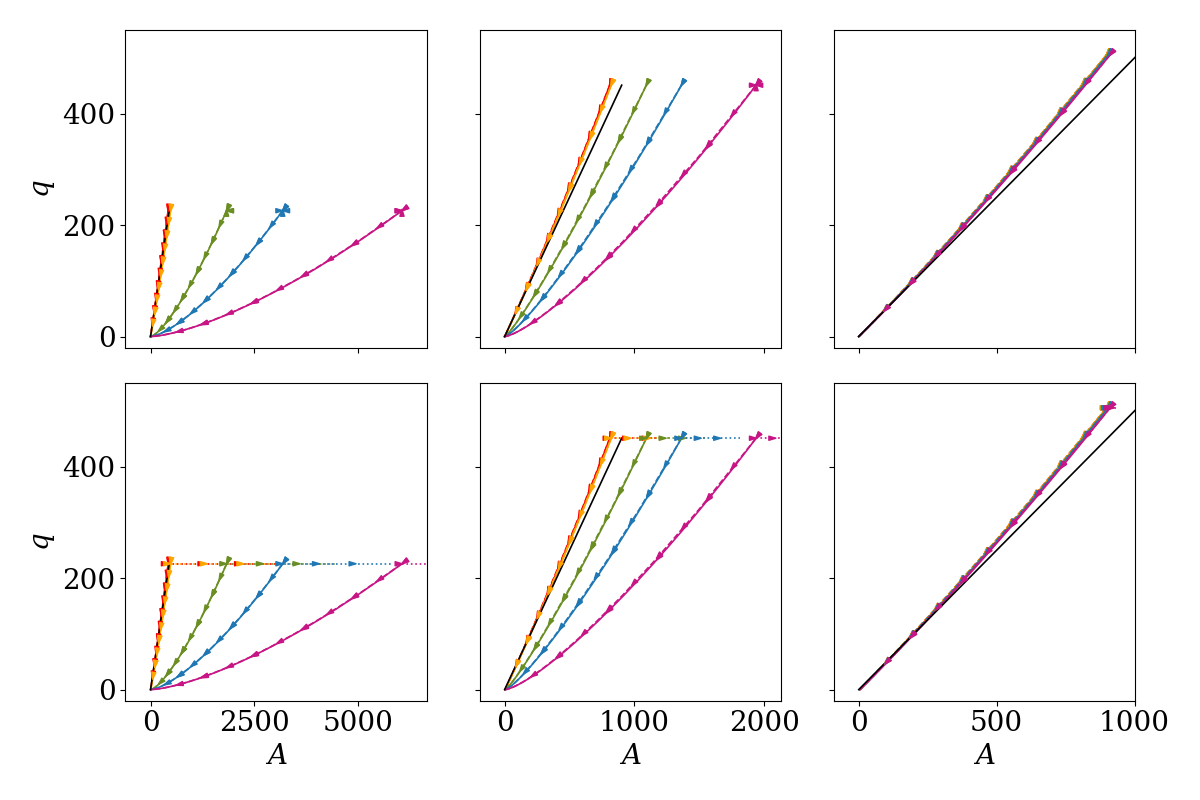}
			\caption{Trajectory examples in $(A,q)$ space for $m_\chi/\omega_0=0$ (top row) and $m_\chi/\omega_0=2$ (bottom row) for second quartic preheating.
				The black $A=2q$ line roughly separates the unstable region above from the stable region below.
				The solid lines correspond to the mixed case $m\neq0,\ \lambda\neq0$, the dashed lines to purely quartic $m=0,\ \lambda\neq0$, and the dotted lines to purely massive $m=0,\ \lambda\neq0$.
				The three columns correspond from left to right to $u_0\equiv \lambda\Phi_0^2/m^2=1, 10, 1000$ in the mixed case, and the parameters for the massive and quartic cases are picked to match $(A_0,q_0)$ of the mixed cases.
				The colors correspond to different wavenumbers $k_0/\omega_0$, listed in Figure \ref{fig:g-mus}.
			}
			\label{fig:g-tjs}
		\end{figure}[t]
		\begin{figure}
			\includegraphics[width=0.7\textwidth]{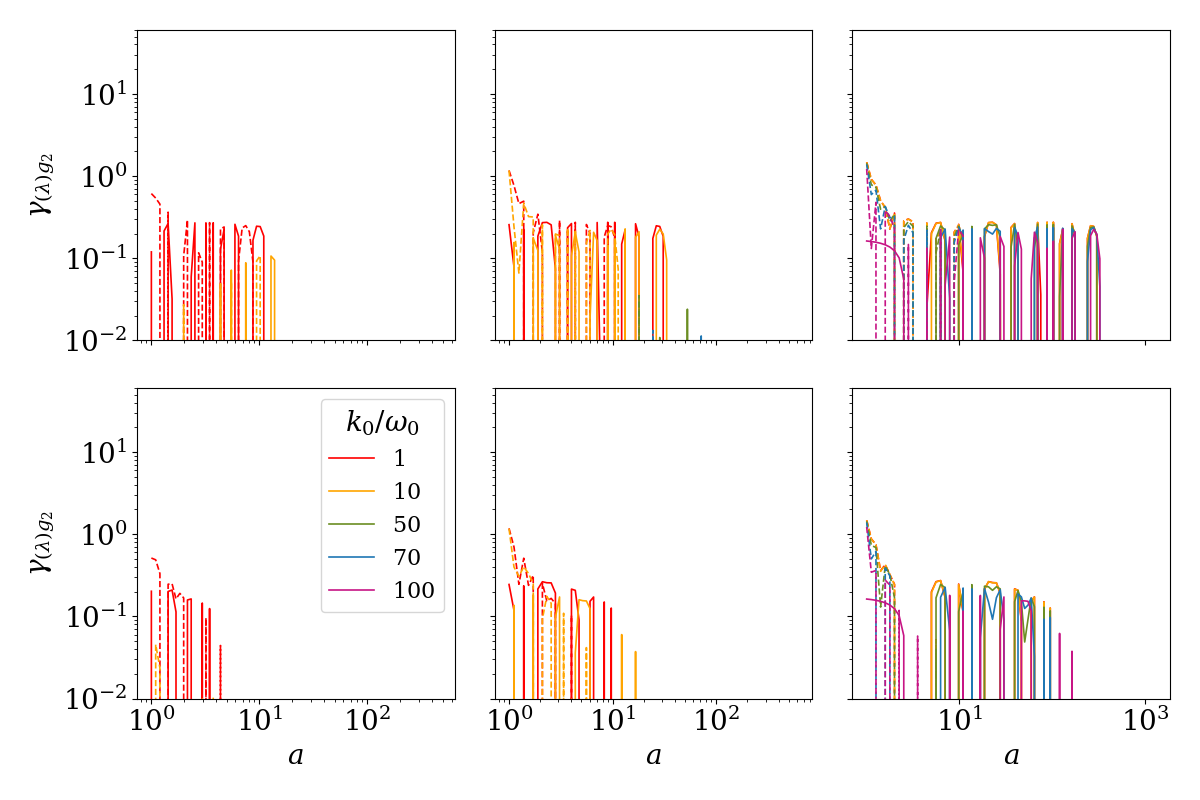}
			\caption{
				Growth rates $\gamma$ per oscillation time along trajectories as a function of scale factor $a$ corresponding to the trajectories in Figure \ref{fig:g-tjs}.
				The colors correspond to different wavenumbers $k_0/\omega_0$; note that for the massive case, $\omega_0=m$, for the quartic case, $\omega_0\approx 0.85\sqrt{\lambda}\Phi_0$, and for the mixed case, $\omega$ is the function shown in equation (\ref{eqn:omega}).
				These growth rates are lower than those from trilinear preheating in Figure \ref{fig:s-mus}, even with similar or slightly greater $q_\mathrm{max}$.}
			\label{fig:g-mus}
		\end{figure}
		We recover a form of Hill's equation with the same $\omega$ and $\epsilon$ expressions as in the trilinear equation of motion, different expressions for $A$ and $q$, and an $\epsilon$-correction proportional to cosine rather than a squared sine.
		As a result, the stability map  in Figure \ref{fig:quartic-g2-stability} looks very similar to the trilinear case.
		
		The trajectories in $(A,q)$ space for the second quartic coupling tend to be farther in the stable regions compared to trajectories for the trilinear case with the same $(A_0,q_0)$, resulting in subdominant growth.
		The trajectories and growth rates are shown in Figures \ref{fig:g-tjs} and \ref{fig:g-mus}.
		Notably, growth rates from the trilinear and second quartic interactions can be comparable when $V(\phi)\not\supset \lambda\phi^4/4$ depending on choices of the $q$ parameters in each \cite{Dufaux_2006};
		trilinear growth remains dominant with our choice of setting ${q_\sigma}_\mathrm{max}\approx{q_{g_2}}_\mathrm{max}$.
		
		Notice that for $m=m_\chi=0$, the pure quartic potential scenario, $(A,q)=(A_0,q_0)=$ {\sl constant} because conformal symmetry is \textit{mostly} preserved, as was shown in \cite{Greene_1997}.
		Following the idea in Appendix \ref{sec:gammaH}, we note that while $\gamma$ is constant for this potential, $\gamma/H$ decreases with time, another reason for the favorable efficiency of the trilinear interaction, for which we showed $\gamma/H$ increases with time.
		
		\subsection{Explosive growth from third quartic coupling}
		In contrast to the behavior of the first quartic coupling, the potential
		\begin{align}
			V(\phi,\chi) =\frac{m^2}{2}\phi^2 + \frac{\lambda}{4}\phi^4 + \frac{m_\chi^2}{2}\chi^2 + g_3 \phi\chi^3
		\end{align}
		results in double exponential growth.
		We can see this in the equation of motion,
		\begin{align}
			0=\ddot \chi_k+\left(k^2+m_\chi^2 + 3g_3\bar\Phi\phi_R\chi_k\right)\chi_k
		\end{align}
		where the $q$-parameter increases by $q\to q_\mathrm{eff}=q\chi_k(t+T)/\chi_k(t)\approx e^\gamma q$ every cycle, so the degree of exponential instability itself increases exponentially.
		This is desirable for efficient preheating.
		However, these extreme growth rate increases necessitate the inclusion of backreaction and rescattering effects to correctly calculate, which is outside the scope of and a reasonable next step from this paper.

		\clearpage
		\section*{Acknowledgements}
		We thank Nemanja Kaloper for useful conversations.  S.W. and J.T.G. thank the Simons Center for hospitality. This research was supported in part by DOE grant DE-FG02-85ER40237. A.B. was partially supported by NSF grant  DMS-2103026, and AFOSR grants FA
		9550-22-1-0215 and FA 9550-23-1-0400. L.B. and F.C.A. also thank the Leinweber Center for Theoretical Physics at the University of Michigan. J.T.G. is supported in part by the National Science Foundation, PHY-2309919. We also thank the referee for useful suggestions on improving the first draft. 
		We'd like to thank our reviewers for constructive comments that improved our thoroughness and quality.

		\bibliography{shortref.bib}

\begin{thebibliography}{33}
\expandafter\ifx\csname natexlab\endcsname\relax\def\natexlab#1{#1}\fi
\expandafter\ifx\csname bibnamefont\endcsname\relax
  \def\bibnamefont#1{#1}\fi
\expandafter\ifx\csname bibfnamefont\endcsname\relax
  \def\bibfnamefont#1{#1}\fi
\expandafter\ifx\csname citenamefont\endcsname\relax
  \def\citenamefont#1{#1}\fi
\expandafter\ifx\csname url\endcsname\relax
  \def\url#1{\texttt{#1}}\fi
\expandafter\ifx\csname urlprefix\endcsname\relax\def\urlprefix{URL }\fi
\providecommand{\bibinfo}[2]{#2}
\providecommand{\eprint}[2][]{\url{#2}}

\bibitem[{\citenamefont{{Guth}}(1981)}]{Guth1981}
\bibinfo{author}{\bibfnamefont{A.~H.} \bibnamefont{{Guth}}},
  \bibinfo{journal}{\prd} \textbf{\bibinfo{volume}{23}}, \bibinfo{pages}{347}
  (\bibinfo{year}{1981}).

\bibitem[{\citenamefont{Amin et~al.}(2014)\citenamefont{Amin, Hertzberg,
  Kaiser, and Karouby}}]{Amin:2014eta}
\bibinfo{author}{\bibfnamefont{M.~A.} \bibnamefont{Amin}},
  \bibinfo{author}{\bibfnamefont{M.~P.} \bibnamefont{Hertzberg}},
  \bibinfo{author}{\bibfnamefont{D.~I.} \bibnamefont{Kaiser}},
  \bibnamefont{and} \bibinfo{author}{\bibfnamefont{J.}~\bibnamefont{Karouby}},
  \bibinfo{journal}{Int. J. Mod. Phys. D} \textbf{\bibinfo{volume}{24}},
  \bibinfo{pages}{1530003} (\bibinfo{year}{2014}), \eprint{1410.3808}.

\bibitem[{\citenamefont{{Lozanov}}(2019)}]{Lozanov2019}
\bibinfo{author}{\bibfnamefont{K.~D.} \bibnamefont{{Lozanov}}},
  \bibinfo{journal}{arXiv e-prints} \bibinfo{eid}{arXiv:1907.04402}
  (\bibinfo{year}{2019}), \eprint{1907.04402}.

\bibitem[{\citenamefont{Greene et~al.}(1997)\citenamefont{Greene, Kofman,
  Linde, and Starobinsky}}]{Greene_1997}
\bibinfo{author}{\bibfnamefont{P.~B.} \bibnamefont{Greene}},
  \bibinfo{author}{\bibfnamefont{L.}~\bibnamefont{Kofman}},
  \bibinfo{author}{\bibfnamefont{A.}~\bibnamefont{Linde}}, \bibnamefont{and}
  \bibinfo{author}{\bibfnamefont{A.~A.} \bibnamefont{Starobinsky}},
  \bibinfo{journal}{Phys. Rev. D} \textbf{\bibinfo{volume}{56}},
  \bibinfo{pages}{6175} (\bibinfo{year}{1997}),
  \urlprefix\url{https://link.aps.org/doi/10.1103/PhysRevD.56.6175}.

\bibitem[{\citenamefont{{Mathieu}}(1868)}]{Mathieu1868}
\bibinfo{author}{\bibfnamefont{E.}~\bibnamefont{{Mathieu}}},
  \bibinfo{journal}{J. Math. Pures Appl.} \textbf{\bibinfo{volume}{13}},
  \bibinfo{pages}{137} (\bibinfo{year}{1868}).

\bibitem[{\citenamefont{{Hill}}(1886)}]{Hill1886}
\bibinfo{author}{\bibfnamefont{G.~W.} \bibnamefont{{Hill}}},
  \bibinfo{journal}{Acta. Math.} \textbf{\bibinfo{volume}{8}},
  \bibinfo{pages}{1} (\bibinfo{year}{1886}).

\bibitem[{\citenamefont{{Magnus} and {Winkler}}(1966)}]{MagWink1966}
\bibinfo{author}{\bibfnamefont{W.}~\bibnamefont{{Magnus}}} \bibnamefont{and}
  \bibinfo{author}{\bibfnamefont{S.}~\bibnamefont{{Winkler}}},
  \emph{\bibinfo{title}{{Hill's Equation}}}, vol.~\bibinfo{volume}{xx}
  (\bibinfo{publisher}{Wiley}, \bibinfo{address}{New York},
  \bibinfo{year}{1966}).

\bibitem[{\citenamefont{\"Ozsoy et~al.}(2017)\citenamefont{\"Ozsoy, Giblin,
  Nesbit, \c{S}eng\"or, and Watson}}]{Ozsoy:2017mqc}
\bibinfo{author}{\bibfnamefont{O.}~\bibnamefont{\"Ozsoy}},
  \bibinfo{author}{\bibfnamefont{J.~T.} \bibnamefont{Giblin}},
  \bibinfo{author}{\bibfnamefont{E.}~\bibnamefont{Nesbit}},
  \bibinfo{author}{\bibfnamefont{G.}~\bibnamefont{\c{S}eng\"or}},
  \bibnamefont{and} \bibinfo{author}{\bibfnamefont{S.}~\bibnamefont{Watson}},
  \bibinfo{journal}{Phys. Rev. D} \textbf{\bibinfo{volume}{96}},
  \bibinfo{pages}{123524} (\bibinfo{year}{2017}), \eprint{1701.01455}.

\bibitem[{\citenamefont{\"Ozsoy et~al.}(2015)\citenamefont{\"Ozsoy, Sengor,
  Sinha, and Watson}}]{Ozsoy:2015rna}
\bibinfo{author}{\bibfnamefont{O.}~\bibnamefont{\"Ozsoy}},
  \bibinfo{author}{\bibfnamefont{G.}~\bibnamefont{Sengor}},
  \bibinfo{author}{\bibfnamefont{K.}~\bibnamefont{Sinha}}, \bibnamefont{and}
  \bibinfo{author}{\bibfnamefont{S.}~\bibnamefont{Watson}}
  (\bibinfo{year}{2015}), \eprint{1507.06651}.

\bibitem[{\citenamefont{Petrov and Blechman}(2016)}]{Petrov:2016azi}
\bibinfo{author}{\bibfnamefont{A.~A.} \bibnamefont{Petrov}} \bibnamefont{and}
  \bibinfo{author}{\bibfnamefont{A.~E.} \bibnamefont{Blechman}},
  \emph{\bibinfo{title}{{Effective Field Theories}}} (\bibinfo{publisher}{WSP},
  \bibinfo{year}{2016}), ISBN \bibinfo{isbn}{978-981-4434-92-8,
  978-981-4434-94-2}.

\bibitem[{\citenamefont{Penco}(2020)}]{Penco:2020kvy}
\bibinfo{author}{\bibfnamefont{R.}~\bibnamefont{Penco}} (\bibinfo{year}{2020}),
  \eprint{2006.16285}.

\bibitem[{\citenamefont{Park et~al.}(2010)\citenamefont{Park, Zurek, and
  Watson}}]{Park:2010cw}
\bibinfo{author}{\bibfnamefont{M.}~\bibnamefont{Park}},
  \bibinfo{author}{\bibfnamefont{K.~M.} \bibnamefont{Zurek}}, \bibnamefont{and}
  \bibinfo{author}{\bibfnamefont{S.}~\bibnamefont{Watson}},
  \bibinfo{journal}{Phys. Rev. D} \textbf{\bibinfo{volume}{81}},
  \bibinfo{pages}{124008} (\bibinfo{year}{2010}), \eprint{1003.1722}.

\bibitem[{\citenamefont{Bloomfield et~al.}(2013)\citenamefont{Bloomfield,
  Flanagan, Park, and Watson}}]{Bloomfield:2012ff}
\bibinfo{author}{\bibfnamefont{J.~K.} \bibnamefont{Bloomfield}},
  \bibinfo{author}{\bibfnamefont{E.~E.} \bibnamefont{Flanagan}},
  \bibinfo{author}{\bibfnamefont{M.}~\bibnamefont{Park}}, \bibnamefont{and}
  \bibinfo{author}{\bibfnamefont{S.}~\bibnamefont{Watson}},
  \bibinfo{journal}{JCAP} \textbf{\bibinfo{volume}{08}}, \bibinfo{pages}{010}
  (\bibinfo{year}{2013}), \eprint{1211.7054}.

\bibitem[{\citenamefont{Gubitosi et~al.}(2013)\citenamefont{Gubitosi, Piazza,
  and Vernizzi}}]{Gubitosi:2012hu}
\bibinfo{author}{\bibfnamefont{G.}~\bibnamefont{Gubitosi}},
  \bibinfo{author}{\bibfnamefont{F.}~\bibnamefont{Piazza}}, \bibnamefont{and}
  \bibinfo{author}{\bibfnamefont{F.}~\bibnamefont{Vernizzi}},
  \bibinfo{journal}{JCAP} \textbf{\bibinfo{volume}{02}}, \bibinfo{pages}{032}
  (\bibinfo{year}{2013}), \eprint{1210.0201}.

\bibitem[{\citenamefont{Wen et~al.}(2022)\citenamefont{Wen, Nesbit, Huterer,
  and Watson}}]{Wen:2021bsc}
\bibinfo{author}{\bibfnamefont{Y.}~\bibnamefont{Wen}},
  \bibinfo{author}{\bibfnamefont{E.}~\bibnamefont{Nesbit}},
  \bibinfo{author}{\bibfnamefont{D.}~\bibnamefont{Huterer}}, \bibnamefont{and}
  \bibinfo{author}{\bibfnamefont{S.}~\bibnamefont{Watson}},
  \bibinfo{journal}{JCAP} \textbf{\bibinfo{volume}{03}}, \bibinfo{pages}{042}
  (\bibinfo{year}{2022}), \eprint{2111.02866}.

\bibitem[{\citenamefont{Brauner et~al.}(2022)\citenamefont{Brauner, Hartnoll,
  Kovtun, Liu, Mezei, Nicolis, Penco, Shao, and Son}}]{Brauner:2022rvf}
\bibinfo{author}{\bibfnamefont{T.}~\bibnamefont{Brauner}},
  \bibinfo{author}{\bibfnamefont{S.~A.} \bibnamefont{Hartnoll}},
  \bibinfo{author}{\bibfnamefont{P.}~\bibnamefont{Kovtun}},
  \bibinfo{author}{\bibfnamefont{H.}~\bibnamefont{Liu}},
  \bibinfo{author}{\bibfnamefont{M.}~\bibnamefont{Mezei}},
  \bibinfo{author}{\bibfnamefont{A.}~\bibnamefont{Nicolis}},
  \bibinfo{author}{\bibfnamefont{R.}~\bibnamefont{Penco}},
  \bibinfo{author}{\bibfnamefont{S.-H.} \bibnamefont{Shao}}, \bibnamefont{and}
  \bibinfo{author}{\bibfnamefont{D.~T.} \bibnamefont{Son}}, in
  \emph{\bibinfo{booktitle}{{Snowmass 2021}}} (\bibinfo{year}{2022}),
  \eprint{2203.10110}.

\bibitem[{\citenamefont{Greene and Kofman}(2000)}]{Greene:2000ew}
\bibinfo{author}{\bibfnamefont{P.~B.} \bibnamefont{Greene}} \bibnamefont{and}
  \bibinfo{author}{\bibfnamefont{L.}~\bibnamefont{Kofman}},
  \bibinfo{journal}{Phys. Rev. D} \textbf{\bibinfo{volume}{62}},
  \bibinfo{pages}{123516} (\bibinfo{year}{2000}), \eprint{hep-ph/0003018}.

\bibitem[{\citenamefont{Cheung et~al.}(2008)\citenamefont{Cheung, Creminelli,
  Fitzpatrick, Kaplan, and Senatore}}]{Cheung:2007st}
\bibinfo{author}{\bibfnamefont{C.}~\bibnamefont{Cheung}},
  \bibinfo{author}{\bibfnamefont{P.}~\bibnamefont{Creminelli}},
  \bibinfo{author}{\bibfnamefont{A.~L.} \bibnamefont{Fitzpatrick}},
  \bibinfo{author}{\bibfnamefont{J.}~\bibnamefont{Kaplan}}, \bibnamefont{and}
  \bibinfo{author}{\bibfnamefont{L.}~\bibnamefont{Senatore}},
  \bibinfo{journal}{JHEP} \textbf{\bibinfo{volume}{03}}, \bibinfo{pages}{014}
  (\bibinfo{year}{2008}), \eprint{0709.0293}.

\bibitem[{\citenamefont{Creminelli et~al.}(2006)\citenamefont{Creminelli, Luty,
  Nicolis, and Senatore}}]{Creminelli:2006xe}
\bibinfo{author}{\bibfnamefont{P.}~\bibnamefont{Creminelli}},
  \bibinfo{author}{\bibfnamefont{M.~A.} \bibnamefont{Luty}},
  \bibinfo{author}{\bibfnamefont{A.}~\bibnamefont{Nicolis}}, \bibnamefont{and}
  \bibinfo{author}{\bibfnamefont{L.}~\bibnamefont{Senatore}},
  \bibinfo{journal}{JHEP} \textbf{\bibinfo{volume}{12}}, \bibinfo{pages}{080}
  (\bibinfo{year}{2006}), \eprint{hep-th/0606090}.

\bibitem[{\citenamefont{Weinberg}(2003)}]{Weinberg:2003sw}
\bibinfo{author}{\bibfnamefont{S.}~\bibnamefont{Weinberg}},
  \bibinfo{journal}{Phys. Rev. D} \textbf{\bibinfo{volume}{67}},
  \bibinfo{pages}{123504} (\bibinfo{year}{2003}), \eprint{astro-ph/0302326}.

\bibitem[{\citenamefont{Arkani-Hamed et~al.}(2004)\citenamefont{Arkani-Hamed,
  Cheng, Luty, and Mukohyama}}]{ArkaniHamed:2003uy}
\bibinfo{author}{\bibfnamefont{N.}~\bibnamefont{Arkani-Hamed}},
  \bibinfo{author}{\bibfnamefont{H.-C.} \bibnamefont{Cheng}},
  \bibinfo{author}{\bibfnamefont{M.~A.} \bibnamefont{Luty}}, \bibnamefont{and}
  \bibinfo{author}{\bibfnamefont{S.}~\bibnamefont{Mukohyama}},
  \bibinfo{journal}{JHEP} \textbf{\bibinfo{volume}{05}}, \bibinfo{pages}{074}
  (\bibinfo{year}{2004}), \eprint{hep-th/0312099}.

\bibitem[{\citenamefont{Easther et~al.}(2011)\citenamefont{Easther, Flauger,
  and Gilmore}}]{Easther:2010mr}
\bibinfo{author}{\bibfnamefont{R.}~\bibnamefont{Easther}},
  \bibinfo{author}{\bibfnamefont{R.}~\bibnamefont{Flauger}}, \bibnamefont{and}
  \bibinfo{author}{\bibfnamefont{J.~B.} \bibnamefont{Gilmore}},
  \bibinfo{journal}{JCAP} \textbf{\bibinfo{volume}{04}}, \bibinfo{pages}{027}
  (\bibinfo{year}{2011}), \eprint{1003.3011}.

\bibitem[{\citenamefont{Armendariz-Picon
  et~al.}(2008)\citenamefont{Armendariz-Picon, Trodden, and
  West}}]{ArmendarizPicon:2007iv}
\bibinfo{author}{\bibfnamefont{C.}~\bibnamefont{Armendariz-Picon}},
  \bibinfo{author}{\bibfnamefont{M.}~\bibnamefont{Trodden}}, \bibnamefont{and}
  \bibinfo{author}{\bibfnamefont{E.~J.} \bibnamefont{West}},
  \bibinfo{journal}{JCAP} \textbf{\bibinfo{volume}{04}}, \bibinfo{pages}{036}
  (\bibinfo{year}{2008}), \eprint{0707.2177}.

\bibitem[{\citenamefont{Cuissa and Figueroa}(2019)}]{Cuissa:2018oiw}
\bibinfo{author}{\bibfnamefont{J.~R.~C.} \bibnamefont{Cuissa}}
  \bibnamefont{and} \bibinfo{author}{\bibfnamefont{D.~G.}
  \bibnamefont{Figueroa}}, \bibinfo{journal}{JCAP}
  \textbf{\bibinfo{volume}{06}}, \bibinfo{pages}{002} (\bibinfo{year}{2019}),
  \eprint{1812.03132}.

\bibitem[{\citenamefont{Adshead et~al.}(2015)\citenamefont{Adshead, Giblin,
  Scully, and Sfakianakis}}]{Adshead:2015pva}
\bibinfo{author}{\bibfnamefont{P.}~\bibnamefont{Adshead}},
  \bibinfo{author}{\bibfnamefont{J.~T.} \bibnamefont{Giblin}},
  \bibinfo{author}{\bibfnamefont{T.~R.} \bibnamefont{Scully}},
  \bibnamefont{and} \bibinfo{author}{\bibfnamefont{E.~I.}
  \bibnamefont{Sfakianakis}}, \bibinfo{journal}{JCAP}
  \textbf{\bibinfo{volume}{12}}, \bibinfo{pages}{034} (\bibinfo{year}{2015}),
  \eprint{1502.06506}.

\bibitem[{\citenamefont{Adshead et~al.}(2018)\citenamefont{Adshead, Giblin, and
  Weiner}}]{Adshead:2018doq}
\bibinfo{author}{\bibfnamefont{P.}~\bibnamefont{Adshead}},
  \bibinfo{author}{\bibfnamefont{J.~T.} \bibnamefont{Giblin}},
  \bibnamefont{and} \bibinfo{author}{\bibfnamefont{Z.~J.}
  \bibnamefont{Weiner}}, \bibinfo{journal}{Phys. Rev. D}
  \textbf{\bibinfo{volume}{98}}, \bibinfo{pages}{043525}
  (\bibinfo{year}{2018}), \eprint{1805.04550}.

\bibitem[{\citenamefont{{Weinstein} and {Keller}}(1985)}]{Weinstein}
\bibinfo{author}{\bibfnamefont{M.~I.} \bibnamefont{{Weinstein}}}
  \bibnamefont{and} \bibinfo{author}{\bibfnamefont{J.~B.}
  \bibnamefont{{Keller}}}, \bibinfo{journal}{SIAM J. Appl. Math.}
  \textbf{\bibinfo{volume}{45}}, \bibinfo{pages}{200} (\bibinfo{year}{1985}).

\bibitem[{\citenamefont{Child et~al.}(2013)\citenamefont{Child, Giblin,
  Ribeiro, and Seery}}]{Child_2013}
\bibinfo{author}{\bibfnamefont{H.~L.} \bibnamefont{Child}},
  \bibinfo{author}{\bibfnamefont{J.~T.} \bibnamefont{Giblin}},
  \bibinfo{author}{\bibfnamefont{R.~H.} \bibnamefont{Ribeiro}},
  \bibnamefont{and} \bibinfo{author}{\bibfnamefont{D.}~\bibnamefont{Seery}},
  \bibinfo{journal}{Phys. Rev. Lett.} \textbf{\bibinfo{volume}{111}},
  \bibinfo{pages}{051301} (\bibinfo{year}{2013}),
  \urlprefix\url{https://link.aps.org/doi/10.1103/PhysRevLett.111.051301}.

\bibitem[{\citenamefont{Dufaux et~al.}(2006{\natexlab{a}})\citenamefont{Dufaux,
  Felder, Kofman, Peloso, and Podolsky}}]{Dufaux:2006ee}
\bibinfo{author}{\bibfnamefont{J.~F.} \bibnamefont{Dufaux}},
  \bibinfo{author}{\bibfnamefont{G.~N.} \bibnamefont{Felder}},
  \bibinfo{author}{\bibfnamefont{L.}~\bibnamefont{Kofman}},
  \bibinfo{author}{\bibfnamefont{M.}~\bibnamefont{Peloso}}, \bibnamefont{and}
  \bibinfo{author}{\bibfnamefont{D.}~\bibnamefont{Podolsky}},
  \bibinfo{journal}{JCAP} \textbf{\bibinfo{volume}{07}}, \bibinfo{pages}{006}
  (\bibinfo{year}{2006}{\natexlab{a}}), \eprint{hep-ph/0602144}.

\bibitem[{\citenamefont{Boyanovsky et~al.}(1996)\citenamefont{Boyanovsky,
  de~Vega, Holman, and Salgado}}]{Boyanovsky1996}
\bibinfo{author}{\bibfnamefont{D.}~\bibnamefont{Boyanovsky}},
  \bibinfo{author}{\bibfnamefont{H.~J.} \bibnamefont{de~Vega}},
  \bibinfo{author}{\bibfnamefont{R.}~\bibnamefont{Holman}}, \bibnamefont{and}
  \bibinfo{author}{\bibfnamefont{J.~F.~J.} \bibnamefont{Salgado}},
  \bibinfo{journal}{Phys. Rev. D} \textbf{\bibinfo{volume}{54}},
  \bibinfo{pages}{7570} (\bibinfo{year}{1996}), \eprint{hep-ph/9608205}.

\bibitem[{\citenamefont{Frasca}(2011)}]{Frasca2011}
\bibinfo{author}{\bibfnamefont{M.}~\bibnamefont{Frasca}},
  \bibinfo{journal}{Journal of Nonlinear Mathematical Physics}
  \textbf{\bibinfo{volume}{18}}, \bibinfo{pages}{291} (\bibinfo{year}{2011}),
  \urlprefix\url{https://doi.org/10.1142%2Fs1402925111001441}.

\bibitem[{\citenamefont{{Abramowitz} and {Stegun}}(1972)}]{Abstegun1972}
\bibinfo{author}{\bibfnamefont{M.}~\bibnamefont{{Abramowitz}}}
  \bibnamefont{and} \bibinfo{author}{\bibfnamefont{I.~A.}
  \bibnamefont{{Stegun}}}, \emph{\bibinfo{title}{{Handbook of Mathematical
  Functions}}} (\bibinfo{year}{1972}).

\bibitem[{\citenamefont{Dufaux et~al.}(2006{\natexlab{b}})\citenamefont{Dufaux,
  Felder, Kofman, Peloso, and Podolsky}}]{Dufaux_2006}
\bibinfo{author}{\bibfnamefont{J.~F.} \bibnamefont{Dufaux}},
  \bibinfo{author}{\bibfnamefont{G.~N.} \bibnamefont{Felder}},
  \bibinfo{author}{\bibfnamefont{L.}~\bibnamefont{Kofman}},
  \bibinfo{author}{\bibfnamefont{M.}~\bibnamefont{Peloso}}, \bibnamefont{and}
  \bibinfo{author}{\bibfnamefont{D.}~\bibnamefont{Podolsky}},
  \bibinfo{journal}{Journal of Cosmology and Astroparticle Physics}
  \textbf{\bibinfo{volume}{2006}}, \bibinfo{pages}{006–006}
  (\bibinfo{year}{2006}{\natexlab{b}}), ISSN \bibinfo{issn}{1475-7516},
  \urlprefix\url{http://dx.doi.org/10.1088/1475-7516/2006/07/006}.

\end{thebibliography}

	\end{document}